\documentclass{aa}
\usepackage{graphicx}
\usepackage{color}
\usepackage{graphicx}
\usepackage{lscape}
\usepackage{natbib}
\usepackage[varg]{txfonts}
\usepackage{amssymb}
\usepackage{stmaryrd}
\usepackage{url}
\usepackage{xspace}
\usepackage{color}
\usepackage{siunitx}
\usepackage{textcomp}
\bibpunct{(}{)}{;}{a}{}{,}
 
\newcounter{Rco}
\newcommand{\Ionst}[1]{\setcounter{Rco}{#1}\Roman{Rco}}
\newcommand{\Ion}[2]{\mbox{#1\,{\scriptsize\Ionst{#2}}}}
\newcommand{\Ionw}[3]{\mbox{#1\,{\scriptsize\Ionst{#2}}~$\lambda\,#3$\,\AA}}

\newcommand{\logg}{\mbox{$\log g$}\xspace}
\newcommand{\loggw}[1]{\mbox{$\log g\hspace{-0.5mm} =\hspace{-0.5mm}  #1$}}

\newcommand{\fg}[1]{\mbox{Fig.\,\ref{#1}}}

\newcommand{\se}[1]{\mbox{Sect.\,\ref{#1}}}

\newcommand{\Teff}{\mbox{$T_\mathrm{eff}$}\xspace}

\newcommand{\ebv}{$E_\mathrm{B-V}$\xspace}

\newcommand{\Msol}{$M_\odot$}
\newcommand{\Rsol}{$R_\odot$}

\begin{document}
 
\title{Mysterious, Variable, and Extremely Hot:\\ White Dwarfs Showing Ultra-High Excitation Lines}
\subtitle{I. Photometric Variability}
\titlerunning{White Dwarfs showing Ultra-High Excitation Lines - Photometric Variability}
 
\author{Nicole Reindl\inst{1}
  \and Veronika Schaffenroth\inst{1}
  \and Semih Filiz\inst{1}
  \and Stephan Geier\inst{1}
  \and Ingrid Pelisoli\inst{1,2}
  \and S. O. Kepler\inst{3}
}
 
\institute{Institute for Physics and Astronomy, University of Potsdam,
  Karl-Liebknecht-Str. 24/25, D-14476 Potsdam, Germany\\
  nreindl885@gmail.com
  \and Department of Physics, University of Warwick, Coventry, CV4 7AL, UK
  \and Instituto de F\'isica, Universidade Federal do Rio Grande do Sul,
  91501-900 Porto-Alegre, RS, Brazil
}
 
\date{Received 6 January 2021 / Accepted 3 February 2021}
 
\abstract{About 10\% of all stars exhibit absorption lines of
  ultra-high excited (UHE) metals (e.g. O\,{\sc viii}) in their optical
  spectra when entering the white dwarf cooling sequence. This is something
  that has never been observed in any other astrophysical object, and
  challenges our understanding of the late stages of stellar evolution since
  decades. The recent discovery of a both spectroscopic and photometric
  variable UHE white dwarf led to the
  speculation that the UHE lines might be created in a shock-heated
  circumstellar magnetosphere.}
  {We aim to gain a better understanding of these mysterious objects by
    studying the photometric variability of the whole population of UHE white dwarfs, and
    white dwarfs showing only the He\,{\sc ii} line problem, as both phenomena
    are believed to be connected.}
  {We investigate (multi-band) light curves from several ground- and
    space-based surveys of all 16 currently known UHE white
   dwarfs (including one newly discovered) and eight white dwarfs that show only the He\,{\sc ii} line
   problem.}
  {We find that $75^{+8}_{-13}$\% of the UHE white dwarfs, and
    $75^{+9}_{-19}$\% of the He\,{\sc ii} line problem white dwarfs are significantly
   photometrically variable, with periods ranging from a 0.22\,d to 2.93\,d
   and amplitudes from a few tenth to a few hundredth mag. The high variability
   rate is in stark contrast to the variability rate amongst normal hot white
   dwarfs (we find $9^{+4}_{-2}$\%), marking UHE and He\,{\sc ii} line problem white
   dwarfs as a new class of variable stars. The period
   distribution of our sample agrees with both the orbital period distribution
   of post-common envelope binaries and the rotational period distribution of
   magnetic white dwarfs if we assume that the objects in our sample will
   spin-up as a consequence of further contraction.}
  {We found further evidence that UHE and He\,{\sc ii} line problem white
    dwarfs are indeed related, as concluded from their overlap in the Gaia HRD, similar
    photometric variability rates, light curve shapes and amplitudes, as well as
    period distributions.
    The lack of increasing photometric amplitudes towards longer wavelengths, as
    well as the non-detection of optical emission lines arising from the highly irradiated face of
    a hypothetical secondary in the optical spectra of our stars, makes it seem
    unlikely that an irradiated late type companion is the origin of the
  photometric variability. Instead, we believe that spots on the surfaces of these
  stars and/or geometrical effects of circumstellar material might be responsible.}

\maketitle
 
%
 
\section{Introduction}
\label{sect:intro}
 
White dwarfs are the end products of the vast majority of all stars,
with about 20\% of them being H-deficient. They are observed over a huge
temperature interval, ranging from 250\,000 K \citep{WernerRauch2015}
down to 2\,710 K \citep{Gianninas+2015}. The early stages of white dwarf cooling
occur very rapidly. When a star enters the white dwarf cooling sequence,
it cools down to 65,000 K within less than a million years, while the
cooling phase down to 3\,000 K takes several billions of years
\citep{Althaus+2009, Renedo+2010}. Thus, although about 37\,000 white dwarfs
have been spectroscopically confirmed to this day \citep{Kepler+2019},
only a tiny fraction (less than 1\%) have effective temperatures (\Teff)
above 65,000\,K.\\
These extremely hot white dwarfs cover a large but sparsely populated region in the
Hertzsprung-Russell Diagram (HRD) and represent an important link in
stellar evolution between the (post-)asymptotic giant branch (AGB) stars,
and the bulk of the white dwarfs on the cooling sequence. Several intriguing
physical processes take place during the early stages of white dwarf cooling that mark those
stars as important astronomical tools even beyond stellar evolution studies.
The intense extreme ultraviolet (UV) flood radiated from a very hot white
dwarf can evaporate giant planets. A fraction of the evaporated volatiles may
then be accreted, which could lead to the pollution of the white dwarf atmosphere
\citep{Gaensicke+2019, Schreiber+2019}. Thus, detailed abundance analyses of
hot white dwarfs bear the potential of reconstructing the composition of
exosolar gaseous planets.
Some white dwarfs in the \Teff interval $58\,000-85\,000$\,K were found to
display high abundances of trans-iron group elements (atomic number $Z>29$),
which is thought to be caused by efficient radiative levitation of those
elements \citep{Chayer+2005, Hoyer+2017, Hoyer+2018, Loebling+2020}.
These stars serve as important stellar laboratories to derive atomic data for
highly ionized species of trans-iron elements \citep{Rauch2012, Rauch2014a,
  Rauch2014b, Rauch+2015a, Rauch+2015b, Rauch+2016, Rauch+2017a, Rauch+2017b}.
Hot white dwarfs have also proven to be powerful tools for Galactic
archaeology and cosmology. They are employed to check a dependency of fundamental
constants, i.e., the fine structure constant $\alpha$, with gravity,
\citep{Berengut+2013, Bainbridge+2017, Hu+2020}, to derive the age of the
Galactic halo \citep{Kalirai2012, Kilic+2019} or to derive the properties of weakly
interacting particles via the hot white dwarf luminosity function
\citep{Isern2008, MillerBertolami2014b, MillerBertolami2014a}.\\
 
A particularly baffling phenomenon that takes place at the beginning of the
white dwarf cooling sequence is the presence of (partly very strong) absorption lines of
ultra-high excited (UHE) metals (e.g. \Ion{N}{7}, \Ion{O}{8}) in the optical
spectra of the hottest white dwarfs. The occurrence of
these obscure features requires a dense environment with temperatures of the
order $10^6$\,K, by far exceeding the stellar effective temperature. A
photospheric origin can therefore be ruled out. Since some of the UHE lines
often exhibit an asymmetric profile shape, it was first suggested that those
lines might form in a hot, optically thick stellar wind
\citep{Werner+1995}.
Another peculiarity of these objects is, that all
show the Balmer or \Ion{He}{2} line problem, meaning that their
Balmer\,/\,\Ion{He}{2} lines are unusually deep and broad and cannot
be fitted with any model.
There are also white dwarfs showing only the Balmer\,/\,\Ion{He}{2} line
problem, but no UHE lines. In case of the H-rich (DA-type) white dwarfs
it was found that the Balmer line problem is to some extent due to the
neglect of metal opacities in the models \citep{Werner1996}. But there
are also cases in which the Balmer line problem
persist, even when sophisticated models are used
\citep{Gianninas+2011, Werner+2018a, Werner+2019}. For the He-dominated
(DO-type) white dwarfs showing the \Ion{He}{2} line problem, however,
even by the addition of metal opacities to the models does not help to
overcome this problem. Since the \Ion{He}{2} line problem is -- without
exception -- observed in every UHE white dwarf, a link between these
two phenomena seems very likely \citep{Werner+2004}. It is thought, that
the \Ion{He}{2} line problem objects are related to the UHE white dwarfs
and that the same process is operating in these stars, but failing to
generate the UHE features \citep{Werner+2014}.\\
The Balmer/\Ion{He}{2} line problem makes it also difficult -- if not
impossible -- to derive accurate temperatures, gravities, and spectroscopic
masses. Some objects show weak \Ion{He}{1} lines, that allow to constrain
their \Teff to some degree. High-resolution UV spectroscopy is available
only for three UHE white dwarfs, which were analyzed by \cite{Werner+2018b}.
They found that the \Teff derived by exploiting several ionization balances
of UV metal lines, agree with what can be estimated from the
\Ion{He}{1}/\Ion{He}{2} ionization equilibrium in the optical. In addition
the study revealed that in these object light metals (C, N, O, Si, P, and S)
are found at generally subsolar abundances and heavy elements from the iron
group (Cr, Mn, Fe, Co, Ni) with solar or over solar abundances. This is not
different from other hot white dwarfs and can be understood as a result of
gravitational settling and radiative levitation of elements.
\cite{Werner+2018b} discussed the possibility that the UHE lines might
form in a multicomponent radiatively driven wind that is frictionally
heated. Such winds are expected to occur in a narrow strip in the
\Teff-\logg-diagram (Fig.\,4 in \citealt{Krticka+Kubat2005}), which indeed
overlaps with the region in which the UHE white dwarfs are observed (see
Fig.\,3 in \citealt{Reindl+2014c}).\\
While this strip could explain why the occurrence of UHE features is restricted to
white dwarfs hotter than $\approx65\,000$\,K, the model does not explain why not all
hot white dwarfs located in this region show this phenomenon. In addition,
the frictionally heated wind model, that assumes a spherically-symmetric
wind, fails to explain the photometric and spectroscopic variability of the
UHE white dwarf J01463+3236 discovered by \cite{Reindl+2019}.
They reported for the first time rapid changes of the
equivalent widths (EWs) of the UHE features in the spectra of J01463+3236, which
were found to be correlated to the photometric period of the star ($\approx 0.24$\,d).
Interpreting this period as the rotational period of the star, they
argue that the UHE features are rotationally modulated and stem from a
co-rotating, shock-heated, circumstellar magnetosphere. Furthermore, they
suggested that the cooler parts of the magnetosphere likely constitute an
additional line forming region of the too-broad and too-deep \Ion{He}{2} lines (or
Balmer lines in case of DAs). White dwarfs which lack the UHE lines and only
show the Balmer/\Ion{He}{2} line problem could then be explained by having
cooler magnetospheres with temperatures not high enough to produce UHE lines.
Since this model requires the white dwarfs to be at least weakly magnetic
(meaning that they should have magnetic field strengths above a few hundred to
thousand Gauss), it could also explain why only a fraction of the hottest white
dwarfs shows UHE lines.\\
The UHE phenomenon affects about 10\% of all stars in the
universe when entering the white dwarf cooling sequence, thus a better
understanding of these objects is highly desirable. Here, we aim to study
the properties of the UHE white dwarfs, as well as their relatives
-- white dwarfs showing only the He\,{\sc ii} line problem -- as a whole. In
particular, we desire to find out if the photometric and
spectroscopic variability observed in J0146+3236 is something that affects all
UHE white dwarfs, and possibly also the He\,{\sc ii} line problem white dwarfs.
This article is the first part of a series of papers and introduces the sample
of UHE and \Ion{He}{2} line problem white dwarfs and investigates their
photometric variability. In \se{sect:sample} we first present the sample
and discuss the location of these stars in the Gaia HRD. Then we search for
photometric variability using light curves from various ground- and space-based
surveys (\se{sect:lightcurves}). The overall results of this study are
presented in \se{sect:results}. Finally, we will discuss our findings
(\se{sect:discussion}) and give an outlook on how more progress can be made
(\se{sect:conclusions}).
 
\section{The sample of UHE and He\,{\sc ii} line problem white dwarfs}
\label{sect:sample}

\begin{table*}
\setlength{\tabcolsep}{0.4em}
\begin{center}
\caption{Names, spectral types, J2000 coordinates, observed Gaia eDR3 $G$ band
  magnitudes, distances, Gaia extinction coefficients, dereddened Gaia color indexes, and the absolute dereddened
$G$ band magnitudes of all known UHE white dwarfs and white dwarfs showing only the He II line problem.}
\label{tab:sample}
\begin{tabular}{l r r r r r c c c}
\hline\hline
\noalign{\smallskip}
Name & Spectral & RA & DEC & $G$ & d  & $A_G$  & $(BP-RP)_0$ &$M_{G_0}$  \\
     & type     & J2000 & J2000   &[mag]  & [pc] &  [mag] & [mag] & [mag] \\
\hline
\noalign{\smallskip}
\multicolumn{6}{l}{UHE white dwarfs}\\
\noalign{\smallskip}
\hline
\noalign{\smallskip}                    
\object{SDSSJ003213.14+160434.8}$^{(a)}$ & DOZ*V UHE  &  8.05472    & 16.07633 & 15.75 & $ 413^{+9}_{-9}$     & 0.13 & -0.63 & 7.55 \\  
\noalign{\smallskip}
\object{WD0101$-$182}$^{(b)}$            & DOZ*V UHE  & 16.06273         & -18.02916 & 15.74 & $398^{+10}_{-9}$ & 0.04 & -0.61 & 7.71 \\
\noalign{\smallskip}
\object{SDSSJ014636.73+323614.3}$^{(c)}$ & DO*V  UHE  &  26.65308  & 32.60403 & 15.54 & $ 331^{+7}_{-6}$      & 0.13 & -0.67 & 7.82 \\
\noalign{\smallskip}
\object{HS\,0158+2335}$^{(d,e)}$         & DO*V  UHE  &  30.36338  & 23.83134 & 16.97 & $ 476^{+17}_{-16 }$  & 0.22 & -0.54 & 8.39 \\
\noalign{\smallskip}
\object{SDSSJ025403.75+005854.5}$^{(f)}$ & DO*V  UHE  &  43.51563  & 0.98173  & 17.41 & $ 764^{+76}_{-63 }$  & 0.18 & -0.59 & 7.87 \\
\noalign{\smallskip}
\object{HE\,0504$-$2408}$^{(g,h)}$       & DO  UHE  &  76.57540  & -24.06685& 15.69 & $ 468^{+11}_{-11 }$     & 0.03 & -0.61 & 7.31 \\        
\noalign{\smallskip}
\object{HS\,0713+3958}$^{(e,g,h)}$       & DO*V  UHE  &   109.26134 & 39.88989 & 16.56 & $ 654^{+35}_{-32 }$  & 0.12 & -0.56 & 7.40 \\    
\noalign{\smallskip}
\object{HS\,0727+6003}$^{(d,h)}$         & DO*V  UHE  &  112.83912 & 59.96028 & 16.09 & $ 426^{+11}_{-11 }$  & 0.13 & -0.62 & 7.83 \\    
\noalign{\smallskip}
\object{HS\,0742+6520}$^{(e)}$           & DO  UHE  &  116.85481 & 65.21699 & 15.73 & $ 332^{+5}_{-5}$      & 0.07 & -0.63 & 8.07 \\          
\noalign{\smallskip}
\object{SDSSJ090023.89+234353.2}$^{(a)}$ & DA  UHE  &  135.09954 & 23.73146 & 18.74 & $2133^{+2675}_{-763}$& 0.06 & -0.62 & 7.29 \\
\noalign{\smallskip}
\object{SDSSJ105956.00+404332.4}$^{(i)}$ & DOZ*V UHE  & 164.98336 & 40.72568 & 18.31 & $2499^{+2391}_{-821}$       & 0.03 & -0.67 & 6.63 \\
\noalign{\smallskip}
\object{SDSSJ121523.08+120300.7}$^{(f)}$ & DOZ*V UHE  &  183.84619 & 12.05022 & 18.14 & $1402^{+349}_{-233}$         & 0.06 & -0.71 & 7.51 \\
\noalign{\smallskip}
\object{SDSSJ125724.04+422054.2}$^{(a)}$ & DA*V  UHE  &  194.35026 & 42.34845 & 17.44 & $ 889^{+96}_{-79 }$      & 0.04 & -0.42 & 7.75 \\
\noalign{\smallskip}
\object{SDSSJ151026.48+610656.9}$^{(f)}$        & DO*V  UHE  & 227.61031 & 61.11581 & 17.26 & $ 786^{+40}_{-36 }$      & 0.02 & -0.59 & 7.84 \\
\noalign{\smallskip}
\object{HS\,2027+0651}$^{(d)}$                     & DO*V  UHE  & 307.38544 & 7.01881  & 16.62 & $ 524^{+19}_{-18 }$      & 0.18 & -0.52 & 7.87 \\
\noalign{\smallskip}
\object{HS\,2115+1148}$^{(d,i,h)}$       & DAO*V UHE  & 319.57804 & 12.02558 & 16.44 & $ 523^{+19}_{-18 }$      & 0.13 & -0.60 & 7.74 \\
\noalign{\smallskip}
\hline
\noalign{\smallskip}
\multicolumn{6}{l}{White dwarfs showing only the He II line problem}\\
\noalign{\smallskip}
\hline
\noalign{\smallskip}
\object{SDSSJ082134.59+173919.4}$^{(i)}$        & DOZ*V UHE: &  125.39562 & 17.65539 & 19.01 & $1173^{+517}_{-275}$         & 0.08 & -0.47 & 8.72 \\
\noalign{\smallskip}
\object{SDSSJ082724.44+585851.7}$^{(i)}$        & DO UHE: &  126.85192 & 58.98104 & 17.47 & $ 579^{+29}_{-27 }$      & 0.32 & -0.37 & 8.36 \\        
\noalign{\smallskip}
\object{SDSSJ094722.49+101523.6}$^{(i)}$        & DOZ UHE:   &  146.84374 & 10.25657 & 18.00 & $ 898^{+143}_{-108}$          & 0.05 & -0.59 & 8.29 \\  
\noalign{\smallskip}
\object{SDSSJ102907.31+254008.3}$^{(a)}$ & DO*V  UHE:    &157.28044 & 25.66901 & 17.05 & $ 583^{+30}_{-27}$      & 0.04 & -0.59 & 8.24 \\
\noalign{\smallskip}
\object{HE1314+0018}$^{(j)}$             & DOZ*V     & 199.35303 & 0.04380  & 16.01 & $ 321^{+8}_{-8}$      & 0.06 & -0.60 & 8.42 \\       
\noalign{\smallskip}
\object{SDSSJ151215.72+065156.3}$^{(i)}$  & DOZ*V    & 228.06540 & 6.86566  & 17.22 & $1019^{+122}_{-98}$      & 0.07 & -0.55 & 7.21 \\
\noalign{\smallskip}
\object{HS\,1517+7403}$^{(k)}$  & DOZ*V    & 229.19388 & 73.86848 & 16.63 & $ 774^{+39}_{-35}$      & 0.06 & -0.61 & 7.19 \\
\noalign{\smallskip}
\object{SDSSJ155356.81+483228.6}$^{(f)}$ & DO*V    & 238.48667 & 48.54126 & 18.61 & $1138^{+183}_{-138}$             & 0.04 & -0.55 & 8.43 \\ 
\hline
\noalign{\smallskip}
\end{tabular}
\tablefoot{~\\
\tablefoottext{a}{\cite{Kepler+2019}}
\tablefoottext{b}{This work}
\tablefoottext{c}{\cite{Reindl+2019}}
\tablefoottext{d}{\cite{Dreizler+1995}}
\tablefoottext{e}{\cite{Reindl+2014c}}
\tablefoottext{f}{\cite{Huegelmeyer+2006}}
\tablefoottext{g}{\cite{Werner+1995}}
\tablefoottext{h}{\cite{Werner+2018b}}
\tablefoottext{i}{\cite{Werner+2014}}
\tablefoottext{j}{\cite{Werner+2004}}
\tablefoottext{k}{\cite{DreizlerHeber1998}}
}
\end{center}
\end{table*}
 
The first two UHE white dwarfs, the DO-type white dwarfs HS\,0713+3958 and HE\,0504-2408, were discovered
by \cite{Werner+1995}. Soon afterwards \cite{Dreizler+1995} announced
three more DO-type UHE white dwarfs (HS\,0158+2335, HS\,0727+6003, and
HS\,2027+0651) as well as the first H-rich UHE white dwarf
(HS\,2115+1148), which they found in the Hamburg-Schmidt (HS) survey
\citep{Hagen1995}. The number of UHE white dwarfs increased even more with the
Sloan Digital Sky Survey (SDSS). \cite{Huegelmeyer+2006} reported two
DO-type UHE white dwarfs and one DOZ (PG\,1159) UHE white dwarf from the SDSS DR4. Within the SDSS
DR10 two more DO-type UHE white dwarfs were found \citep{Werner+2014, Reindl+2014c},
and \cite{Kepler+2019} announced the discovery of two more
DA-type UHE white dwarfs as well as one (possibly two) more DO-type UHE white dwarfs within the
SDSS DR14. One more DO-type UHE white dwarf was discovered by
\cite{Reindl+2019} based on spectroscopic follow-up of UV-bright sources.
Finally, we announce the discovery of a 16th member of the UHE white dwarfs,
the DOZ-type WD0101$-$182. In archival UVES spectra of this star
($R\approx18\,500$, ProgID 167.D-0407(A), PI: R. Napiwotzki), we detect
for the first time UHE lines around 3872, 4330, 4655, 4785, 5243, 5280,
6060, 6477\,\AA\ (\fg{fig:spectraUHE}). Using non-LTE models for DO-type
white dwarfs \citep{Reindl+2014c, Reindl+2018a} that were calculated
  with the T{\"u}bingen non-LTE Model-Atmosphere Package (\textsc{TMAP},
  \citealt{werner+2003, tmap2012}), we find that the weak
\Ionw{He}{1}{5876} line and the \Ionw{C}{4}{5803, 5814} doublet are best
reproduced with \Teff$=90\,000$\,K and $C=0.003$ (mass fraction).\\
Besides these 16 UHE white dwarfs, our sample includes eight more objects
which show only the \Ion{He}{2} line problem but no clear sign of UHE lines.
The prototype of this class of stars is HE\,1314+0018 that was discovered
by \cite{Werner+2004}. The high-resolution and high signal to noise
spectrum of HE\,1314+0018 lacks any UHE absorption lines. The other seven objects are from the
samples of \cite{DreizlerHeber1998}, \cite{Werner+2014}, and
\cite{Kepler+2019}. Four of them possibly show the UHE feature around
5430-5480\,\AA, which is also one of
the strongest UHE features observed in the UHE white dwarfs.\\
\begin{figure}[ht]
\centering
\includegraphics[width=\columnwidth]{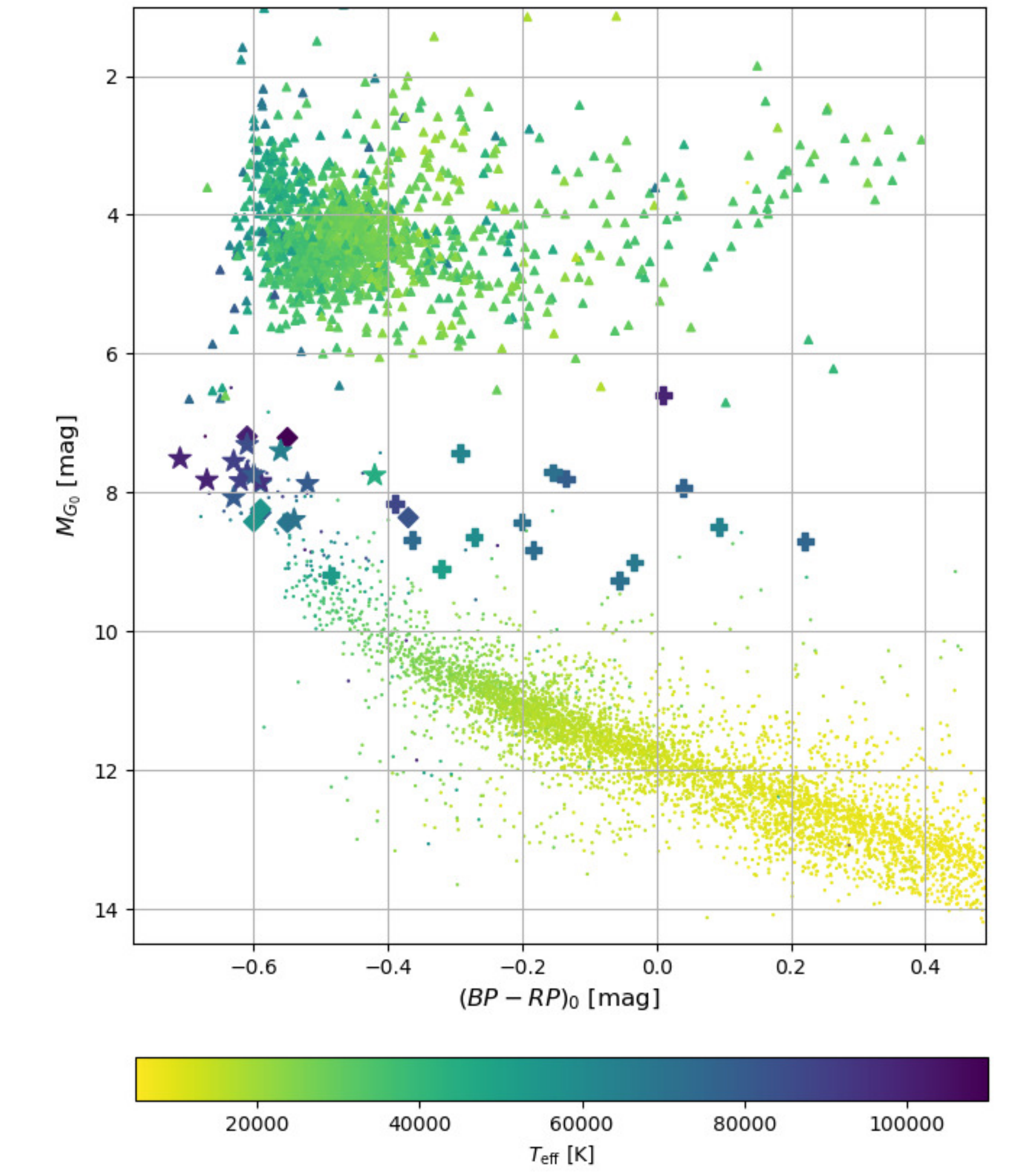}
\caption{Locations of the UHE white dwarfs (star symbols) and white dwarfs
  showing only the He\,{\sc ii} line problem (diamonds) in the Gaia HRD.
  Hot subdwarfs (triangles), SDSS white dwarfs (dots), as well as
  white dwarf-main sequence binaries (plus symbols) containing a very hot
  (\Teff$\ge 50\,000$\,K) white dwarf are also shown.
  The color coding indicates the effective temperatures of the stars.}
\label{fig:gaia}
\end{figure}
In \fg{fig:spectraUHE} and \fg{fig:spectraHeIIP} we show the optical spectra of
all UHE white dwarfs and spectra of all white dwarfs showing only the
\Ion{He}{2} line problem, respectively.
For HS2027+0651 HST/STIS spectra are shown that were
  observed with the G430L and G750L gratings ($R\approx700$).
  We downloaded these observations from the Mikulski Archive for
Space Telescopes (MAST, proposal IDs:
8422, 7809, PIs: H. Ferguson and C. Leitherer, respectively).
For WD0101-182, the UVES spectrum (see above), and for
HE0504-2408 an EFOSC\,1 spectrum obtained at the
ESO 3.6\,m telescope ($R\approx1500$, \citealt{Werner+1995}) are
shown. The spectra of J0146+3236, HS0158+2335, HS0713+3958, HS0727+6003,
HS0742+6520, and HE\,1314+0018 were obtained by us in
October/November 2014 at the Calar Alto $3.5$\,m telescope
(ProgID H14-3.5-022, see also \citealt{Reindl+2019}). We used
the TWIN spectrograph and a slit width of 1.2 acrsec.
For the blue channel grating No. T08, and for the red channel grating
No. T04 were used. The spectra have a
resolution of $1.8\,\AA$. After each spectrum, we required
ThAr wavelength calibration. The data reduction was done
using IRAF. We did not flux-calibrate our data.
For HS1517+7403 and HS2115+1148 TWIN spectra are shown
that were obtained by \cite{Dreizler+1995} and
\cite{DreizlerHeber1998} and which have a resolution of
$3.5\,\AA$. For the remaining objects SDSS spectra ($R\approx1800$)
are shown. Overplotted in red are \textsc{TMAP} models with atmospheric
parameters determined within this work (WD0101$-$182) or with
parameters reported by previous works (see footnote of
Table~\ref{tab:sample}).\\
 
Table~\ref{tab:sample} lists all UHE and \Ion{He}{2} line problem
white dwarfs along with their spectral
types, J2000 coordinates, observed Gaia early DR3 $G$ band magnitudes
\citep{Gaia+2016, Gaia+2018},
distances, $d$, Gaia extinction coefficients, $A_G$,
the dereddened Gaia color indexes, $(BP-RP)_0$, and the absolute dereddened
$G$ band magnitudes.
A spectral type DOZ UHE indicates a He-rich white dwarf that shows photospheric
metal lines in the optical as well as UHE lines. A spectral subtype UHE:
indicates an object with an uncertain identification of UHE lines.
The distances have been calculated from the parallaxes (via $1/\pi$),
which we corrected for the zeropoint bias using the Python code provided by
\cite{Lindegren+2020}\footnote{\url{https://gitlab.com/icc-ub/public/gaiadr3_zeropoint}}.
Following \cite{GentileFusillo+2019}, we assume that the extinction
coefficient $A_G$ in the Gaia $G$ passband scales as $0.835\times A_V$ based
on the nominal wavelengths of the respective filters and the reddening versus
wavelength dependence employed by \cite{Schlafly2011}. Values for $A_V$ were
obtained from the 3D dust map of \cite{Lallement+2018} using the distance
calculated from the Gaia parallax of each object. Nine of our stars are located outside of
the \cite{Lallement+2018} 3D dust map (that is stars with a distance from the
Galactic plane of $|z|\gtrsim 500$\,pc). For those we obtained $A_V$ from the 2D dust
map of \cite{Schlafly2011} and assumed that $A_G$ scales with a factor of
$1-exp(-|z|/200\mathrm{pc})$, as most of the absorbing
material along the line of sight is concentrated
along the plane of the Galactic disk. We note, that the difference in
reddening obtained from the two methods varies by a factor of 0.65 to 2.24 for stars
located within the 3D dust map ($-500\,\mathrm{pc}<z<500\,\mathrm{pc}$). This
demonstrates that an accurate determination is not easy.
The color indies, $(BP-RP)_0$ were calculated using Eq. 18 and
19 in \cite{GentileFusillo+2019}. The absolute Gaia magnitude in the G band
was calculated via $M_{G_0} =G - A_G+ 5 + 5\times\,\log(1/\pi)$, where $\pi$ is
the zero point corrected parallax in milli arcsec from the Gaia early DR3.\\
In Fig.~\ref{fig:gaia} we show the locations of the UHE white dwarfs
(star symbols) and white dwarfs showing only the \Ion{He}{2} line problem
(diamonds) that have parallaxes better than 20\% in the Gaia
HRD. Also shown are the locations of white dwarfs from
the SDSS (dots) with Gaia parallaxes better than 5\,\% and a reddening smaller than
\ebv$<$0.015 \citep{Gaia+2018}, as well as hot subdwarfs (triangles) from
\cite{Geier2020} with Gaia parallaxes better than 20\,\%. The latter were
dereddened following the approach of \cite{GentileFusillo+2019}. Finally,
we also show the locations of white dwarf-main sequence binaries (bold plus
signs) from the sample of \cite{Rebassa-Mansergas+2010} that contain a
very hot (\Teff$\ge 50\,000$\,K) white dwarf primary and have parallaxes
better than 30\,\%.\\
It can be seen that the UHE white dwarfs and white dwarfs showing
only the \Ion{He}{2} line problem overlap in a narrow region
($-0.71$\,mag $\leq (BP-RP)_0 \leq -0.37$\,mag, and 7.19\,mag
$\leq M_G \leq 8.43$\,mag). Both are located well below the hot subdwarf
cloud and are just on top of the white dwarf banana\footnote{The term ``white
dwarf banana'' was coined by \cite{Girven+2011}.}. It also becomes obvious, that
the stars in our sample are amongst the bluest objects. Most of the
hot white dwarfs with an M-type companion are found at similar absolute
magnitudes, but redder colors.
This is a consequence of the flux of the low mass companion that significantly
contributes to the flux in the optical wavelength range. The only object
from the sample of \cite{Rebassa-Mansergas+2010} that directly lies on the
white dwarf banana is \object{SDSS\,J033622.01-000146.7}. For this object
the late type companion is not noticeable in the continuum flux (no increased
flux at longer wavelengths) and also shows no absorption lines from the secondary.
Only the emission lines in the core of the Balmer series are seen, which
originate from the close and highly irradiated side of the cool companion.
Two of our stars, the DA-type UHE white dwarf J1257+4220 and J0827+5858, which
shows only the \Ion{He}{2} line problem, are found at noticeably redder
colors ($-0.42$\,mag and $-0.37$\,mag, respectively) than the rest of our sample.
While J0827+5858 is located at a region with a particularly high reddening
$A_g=0.32$\,mag, which might be underestimated by the 3D dust map, this is unlikely
the case for J1257+4220 ($A_G=0.04$\,mag). Looking at the Gaia eDR3 RUWE
(Renormalized Unit Weight Error) values of our stars, we find they all
have a value close to one (indicating that the single-star model provides a
good fit to the astrometric observations), except for J1257+4220. Here we find
a RUWE value much larger than one, namely 1.3387. This might suggest that J1257+4220
is a (wide) binary or it was otherwise problematic for the astrometric solution.
\\
The mean dereddend color index of our sample is $\overline{BP-RP_0}=-0.58$\,mag
(standard deviation $\sigma=-0.08$\,mag), with the UHE white dwarfs
being slightly bluer ($\overline{BP-RP_0}=-0.60$\,mag, $\sigma=-0.07$\,mag) than white dwarfs
showing only the \Ion{He}{2} line problem ($\overline{BP-RP_0}=-0.54$\,mag, $\sigma=-0.08$\,mag).
We also find that the mean dereddened absolute G band magnitude
of the UHE white dwarfs with parallaxes better than 20\%
($\overline{M_G}=7.76$\,mag, $\sigma=-0.27$\,mag) is slightly brighter than
that of white dwarfs showing only the \Ion{He}{2} line problem
($\overline{M_G}=8.02$\,mag, $\sigma=-0.56$\,mag).\\
We note that 18 out of the 24 stars in our sample have a probability of
being a white dwarf (PWD) greater than 90\% as defined by
\cite{GentileFusillo+2019}. For the remaining objects PWDs
between 72\% and 89\% are found. The only object that is not included in
the catalog of \cite{GentileFusillo+2019} is J0900+2343, which
is also the only object in our sample that had a negative parallax in the
Gaia DR2. For comparison, the catalog of hot subdwarf candidates from the Gaia DR2 by
\cite{Geier+2019} contains only three of our stars. This is because
for objects with parallaxes better than 20\% \cite{Geier+2019} included only
objects with absolute magnitudes between $-1.0\mathrm{mag} \leq M_G \leq 7.0$\,mag.
 
\section{Light curve analysis}
\label{sect:lightcurves}
The discovery of a photometric variability in the UHE white dwarf
J01463+3236 raises the question if photometric variability is a
common feature of UHE white dwarfs, and possibly also of the He\,{\sc ii}
line problem white dwarfs. Here we want to investigate this possibility
by searching for periodic signals in the light curves of these objects.
 
For the analyses of the light curves we used the \textsc{VARTOOLS} program
\citep{HartmanBakos2016} to perform a generalized Lomb-Scargle (LS) search
\citep{ZechmeisterKuerster2009, Press1992} for periodic sinusoidal signals.
We classify objects that show a periodic signal with a false alarm probability
(FAP) of $\log(FAP)\leq-4$ as significantly variable, objects with
$-3\leq\log(FAP)<-4$ as possibly variable, and objects that only show a
periodic signal with $\log(FAP)>-3$ as not variable.
In case we found more than one significant period, we whitened the light
curve by removing the strongest periodic signal (including its harmonics and
subharmonics) from the light curve. Then the periodogram was recomputed
to check if the FAP of the next strongest signal still remains above our
variability threshold ($\log(FAP)\leq-4$). This whitening procedure was repeated
until no more significant periodic signals could be found.\\
Using the -killharm command we fitted a harmonic series of the form
$$m(t) = A\times \sin\left(\frac{2\pi\,(t-t_0)}{P}\right)-B\times\cos\left(\frac{2\pi\,(t-t_0)}{P}\right)+m_0\,\, (1)$$
to each light curve. By that we determine the peak-to-peak amplitude
of the light curve, which we define as the difference between the maximum
and minimum of the fit. The same function was also used to estimate
the uncertainties on the derived periods by running a
Differential Evolution Markov Chain Monte Carlo (DEMCMC) routine
\citep{TerBraa+Cajok2006} employing the -nonlinfit command.
The number of accepted links was set to 10\,000. As initial
guesses we used the period obtained from the LS search, and for the
remaining parameters the values from the killharm fit.\\
In Table~\ref{tab:periods} and Table~\ref{tab:periods2}
we summarize the light curves used in our analysis, data points of each light
curve, mean magnitude in each band, the median value of each period and
its uncertainty as calculated in the DEMCMC simulation, and amplitudes for the UHE
white dwarfs and white dwarfs showing only the \Ion{He}{2} line problem,
respectively. In the following we give now an overview of the data sets
used in our work (\se{sect:data}) and then provide notes on individual objects
(\se{sect:notes}).
 
\subsection{Data sets}
\label{sect:data}
 
Light curves were obtained from various surveys as well as our
own observing campaign.
\begin{figure}[ht]
\centering
\includegraphics[width=0.92\columnwidth]{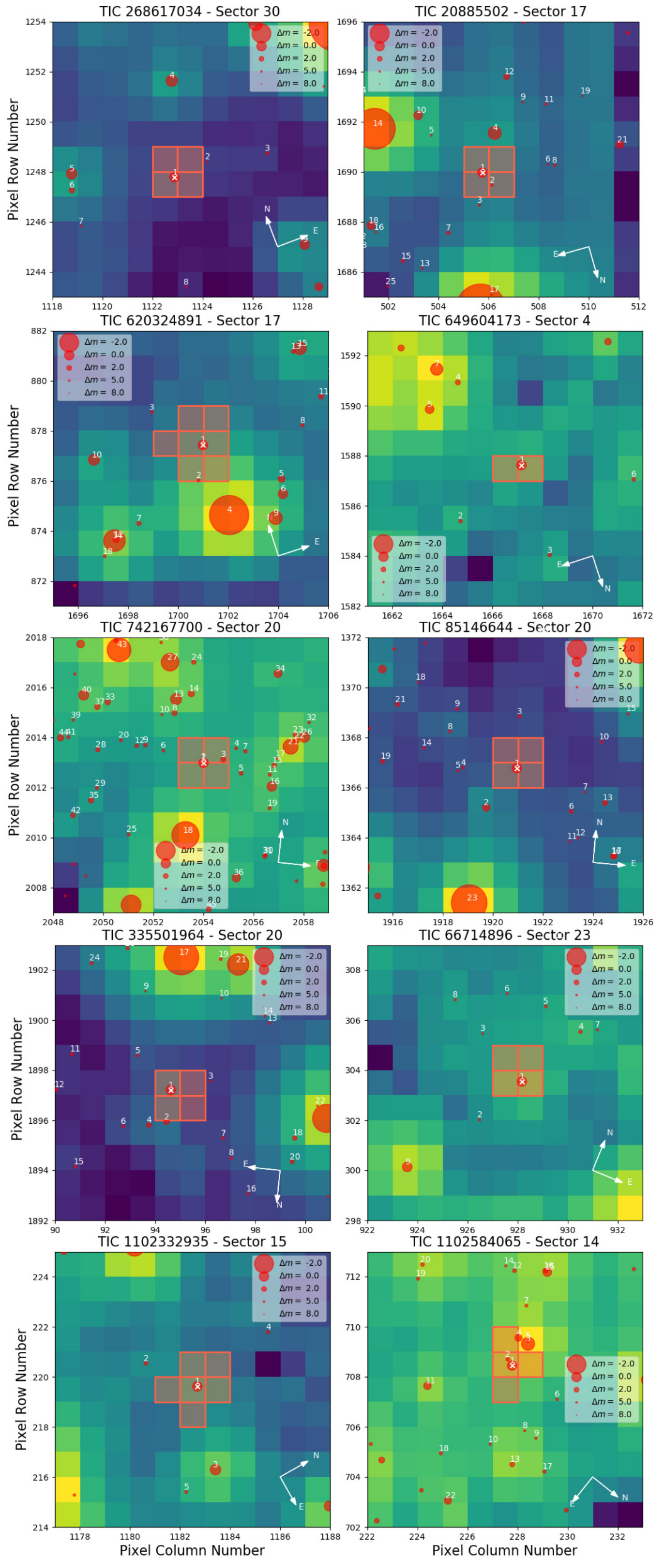}
\caption{From left to right and top to bottom: target pixel files (TPFs) of
  WD0101$-$182, J0146+3236, HS0158+2335, J0254+0058, HS0713+3958, HS0727+6003, HS0742+6520,
  HE1314+0018, J1510+6106, and HS1517+7403. The red
  circles are the sources of the Gaia catalogue in the field with scaled
  magnitudes (see legend). Number 1 indicates the location of the targets.
  The aperture mask used by the pipeline to extract the photometry is also marked.}
\label{fig:tpf}
\end{figure}
\paragraph{TESS}
The Transiting Exoplanet Survey Satellite (TESS) scans the sky with
26 segments and with a 27.4 day observing period per segment.
TESS uses a red-optical bandpass covering the wavelength
range from about 6000 to 10\,000\,\AA\ and which is centered on 7865\,\AA,
like the traditional Cousins I-band.
We downloaded the target pixel files (TPF) of each object from MAST
as FITS format. The FITS files are already processed based
on the  Pre-Search Data Conditioning Pipeline \citep{Jenkins+2016}
from  where  we  have  extracted the barycentric corrected dynamical Julian days
("BJD - 2457000", a  time  system  that  is  corrected  by  leap  seconds,
see  Eastmanet  al.  2010)  and the pre-search Data Conditioning
Simple Aperture Photometry flux ("PDCSAP  FLUX") for which long-term
trends have been removed using the co-trending basis vectors.  In this work,
we  used  the  PDC  light curves and converted the fluxes to fractional
variations from the mean (i.e. differential intensity).
Since TESS has a poor spatial resolution (one detector pixel
corresponds to 21\,arcsec on the sky) and our targets are faint, we carefully
checked for blends with close by stars using the tpfplotter code
\citep{Aller+2020}.
In \fg{fig:tpf} we show the TPF plots for the UHE and \Ion{He}{2} line problem white
dwarfs. The red circles represent Gaia sources, which are scaled by
magnitude contrast against the target source. Also shown is the aperture
mask used by the pipeline to extract the photometry. In total, ten UHE, and
two \Ion{He}{2} problem white dwarfs were observed by TESS in the
two-minute cadence mode.
\paragraph{K2}
In a series of sequential observing campaigns 20 fields, which were
distributed around the ecliptic plane, were observed by the K2 mission
(campaign  duration $\approx80$\,d, \citealt{Howell+2014}).
Throughout the mission K2 observed in two cadence modes,
long cadence ($\approx30$\,min data-point cadence) and short cadence
($\approx1$\,min data-point cadence). The latter was only provided
for selected targets, and the long cadence was used as the default
observing mode. Two of the stars in our sample, J0821+1739 and
J0900+2343, were observed in the long cadence mode.
K2 data contain larger systematics than the original Kepler mission.
This is because of the reduction in pointing precision as a result
of the spacecraft drift during the mission. Thus, several pipelines
have been developed to process K2 light curves. Here, we are using
the light curves produced by the K2 Self Flat Fielding (K2SFF,
\citealt{Vanderburg+Johnson2014}) and the EPIC Variability Extraction
and Removal for Exoplanet Science Targets (EVEREST,
\citealt{Lunger+2016, Lunger+2018}) pipelines. The data were obtained
from the MAST archive.
\paragraph{ATLAS}
Since 2015 the Asteroid Terrestrial-impact Last Alert System (ATLAS,
\citealt{Tonry+2018}), surveys about 13,000\,deg$^2$ at least
four times per night using two independent and fully robotic 0.5\,m telescopes
located at Haleakala and Mauna Loa in Hawaii. It provides c- and o-band light curves
(effective wavelengths $0.53\,\mathrm{\mu m}$ and $0.68\,\mathrm{\mu m}$,
respectively) which are taken with an exposure time of 30\,s.
Eight stars in our sample have ATLAS light curves.
\paragraph{Catalina Sky Survey}
The Catalina Sky Survey uses three 1\,m class
telescopes to cover the sky between declination $-75$\textdegree
$< \delta < +65$\textdegree, but avoids the crowded Galactic plane region
by 10 to 15 degrees due to reduced source recovery. It consists of the
Catalina Schmidt Survey (CSS), the Mount Lemmon Survey (MLS) in Tucson,
Arizona, and the Siding Spring Survey (SSS) in Siding Spring, Australia.
The second data release contains V-band photometry for about 500 million
objects with V magnitudes between 11.5
and 21.5 from an area of 33,000 square degrees \citep{Drake+2009, Drake+2014}.
Most of the stars in our sample are covered by this survey, though
we find that at least 200-300 data points are needed for finding
a periodic signal. This is likely because of the larger uncertainties
on the photometric measurements compared to other surveys employed in
this work.
\paragraph{SDSS stripe 82}
The SDSS Stripe 82 covers an area of 300\,deg$^2$ on the Celestial
Equator, and has been repeatedly scanned in the u-, g-, r-, i-, and z-bands
by the SDSS imaging survey \citep{Abazajian+2009}. For J0254+0058, the only object
in our sample that is included in the SDSS stripe 82  we acquired the
u-, g-, r-, i-, and z-band light curves (about 70 data points each)
from \cite{Ivezic+2007}.
\paragraph{ZTF}
The Zwicky Transient Facility (ZTF, \citealt{Bellm+2019, Masci+2019}) survey
uses a 48-inch Schmidt telescope with a 47\,deg$^2$ field of view, which
ensures that the ZTF can scan the entire northern sky every night.
We obtained data from the DR4 which were acquired between March 2018 and
September 2020, covering a time span of around 470 days. The photometry is
provided in the g, r, and -- less frequent -- in the i-band, with a uniform
exposure time of 30\,s per observation. Most objects in our sample are
covered by this survey, with 21 having at least 50 data points in at
least one band.
\paragraph{BUSCA}
For HS\,0727+6003 we obtained
photometry using the Bonn University Simultaneous Camera (BUSCA,
\citealt{Reif+1999}) at the 2.2\,m telescope at the Calar Alto
Observatory. The star was observed during two consecutive nights
on 21 and 22 December 2018.
The beamsplitters of BUSCA allow to collect visible light
simultaneously in four different bands namely $U_{B}$, $B_{B}$, $R_{B}$, and
$I_{B}$. However, due to technical problems with BUSCA, we could not obtain
data in the $I_{B}$ band. Instead of filters, we used the intrinsic
transmission curve given by the beam splitters to avoid light loss.
For the data reduction, IRAF's aperture photometry package was utilized.
 
\begin{figure*}[ht]
\centering
\includegraphics[width=\textwidth]{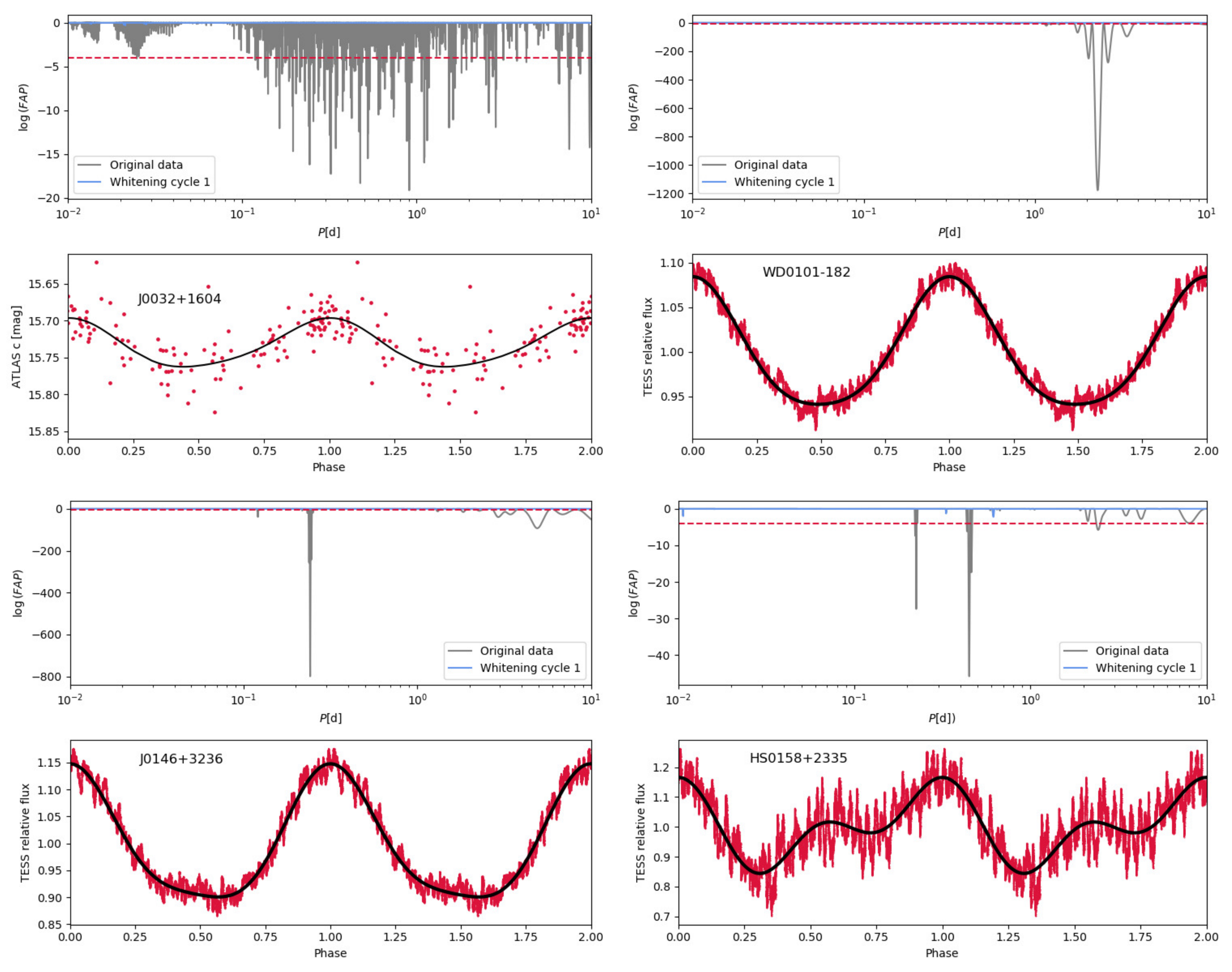}
\caption{Periodograms and phase-folded light curves of  the UHE white dwarfs J0032+1604, WD0101-182,
  J0146+3236, and HS\,0158+2335. The red solid lines are phase-averaged light curves, while the dotted light curve represents the actual
  data. The black line is a fit of a harmonic series used to predict the
peak-to-peak amplitude.}
\label{fig:lp-1}
\end{figure*}
 
\begin{figure*}[ht]
\centering
\includegraphics[width=\textwidth]{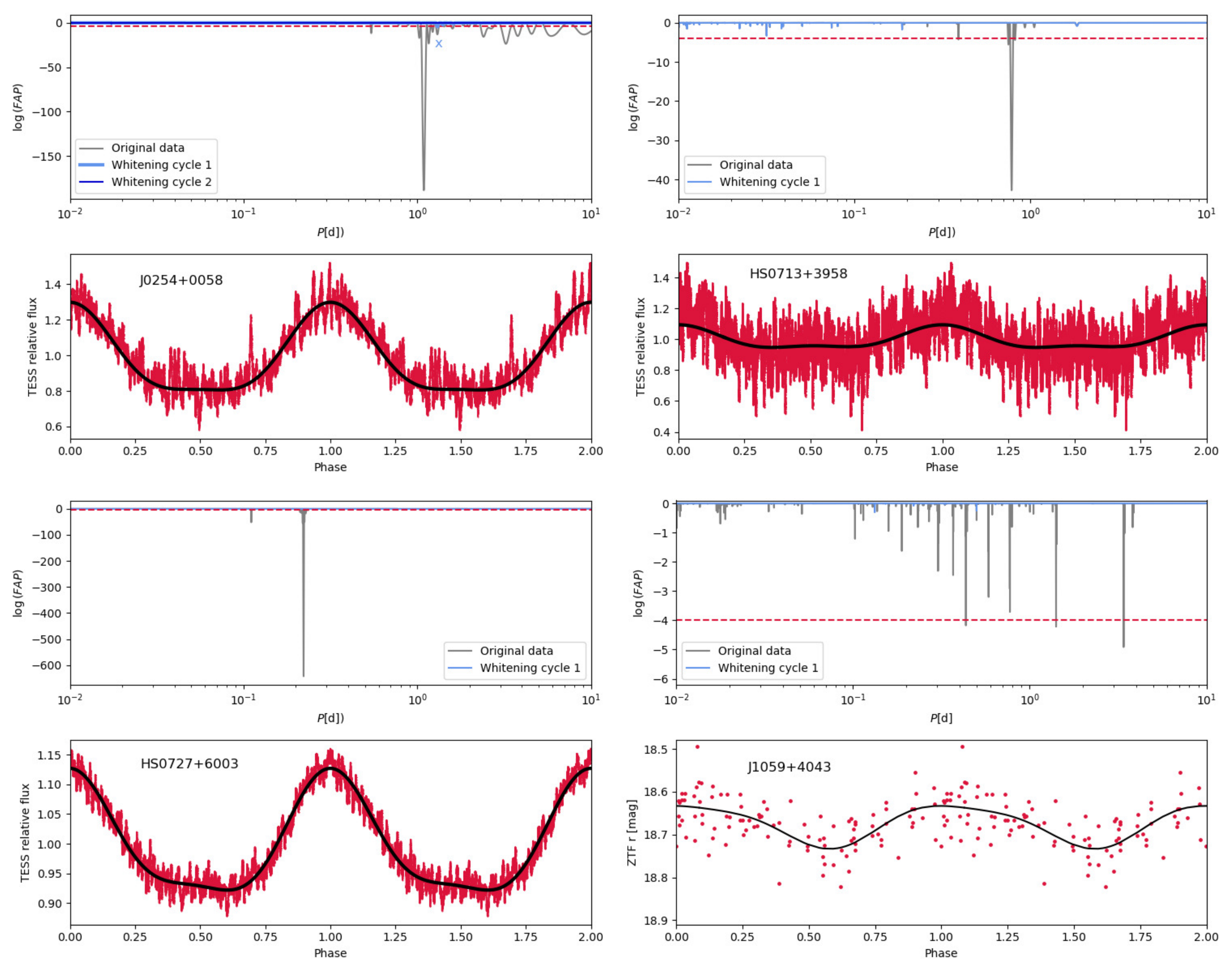}
\caption{Like \fg{fig:lp-1} for the UHE white dwarfs J0254+0058, HS0713+3958, HS0727+6003, and J1059+4043.}
\label{fig:lp-2}
\end{figure*}
 
\begin{figure*}[ht]
\centering
\includegraphics[width=\textwidth]{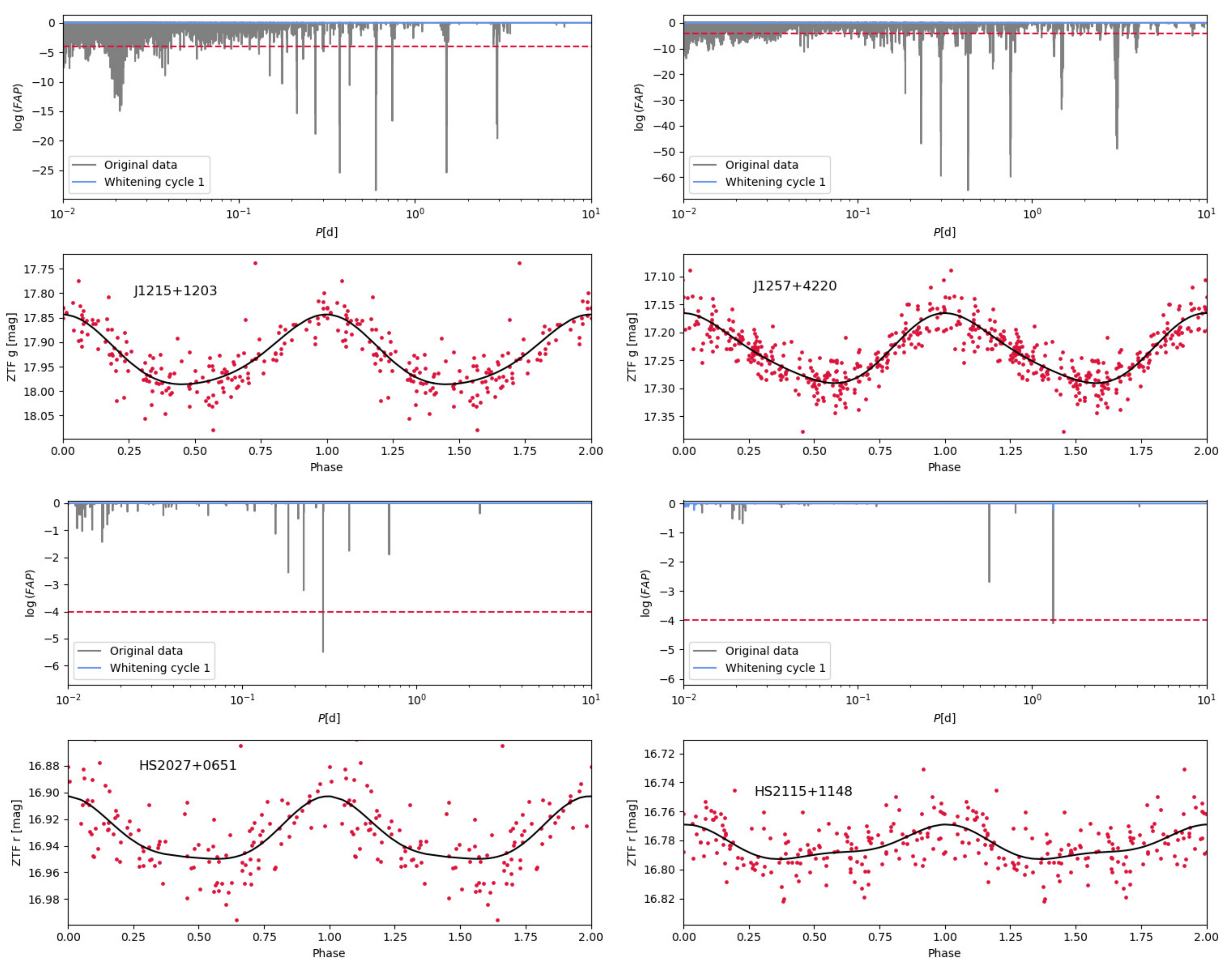}
\caption{Like \fg{fig:lp-1} for the UHE white dwarfs J1215+1203, J1257+4220, HS\,2027+0651, and HS\,2115$-$1148.}
\label{fig:lp-3}
\end{figure*}
 
\begin{figure*}[ht]
\centering
\includegraphics[width=\textwidth]{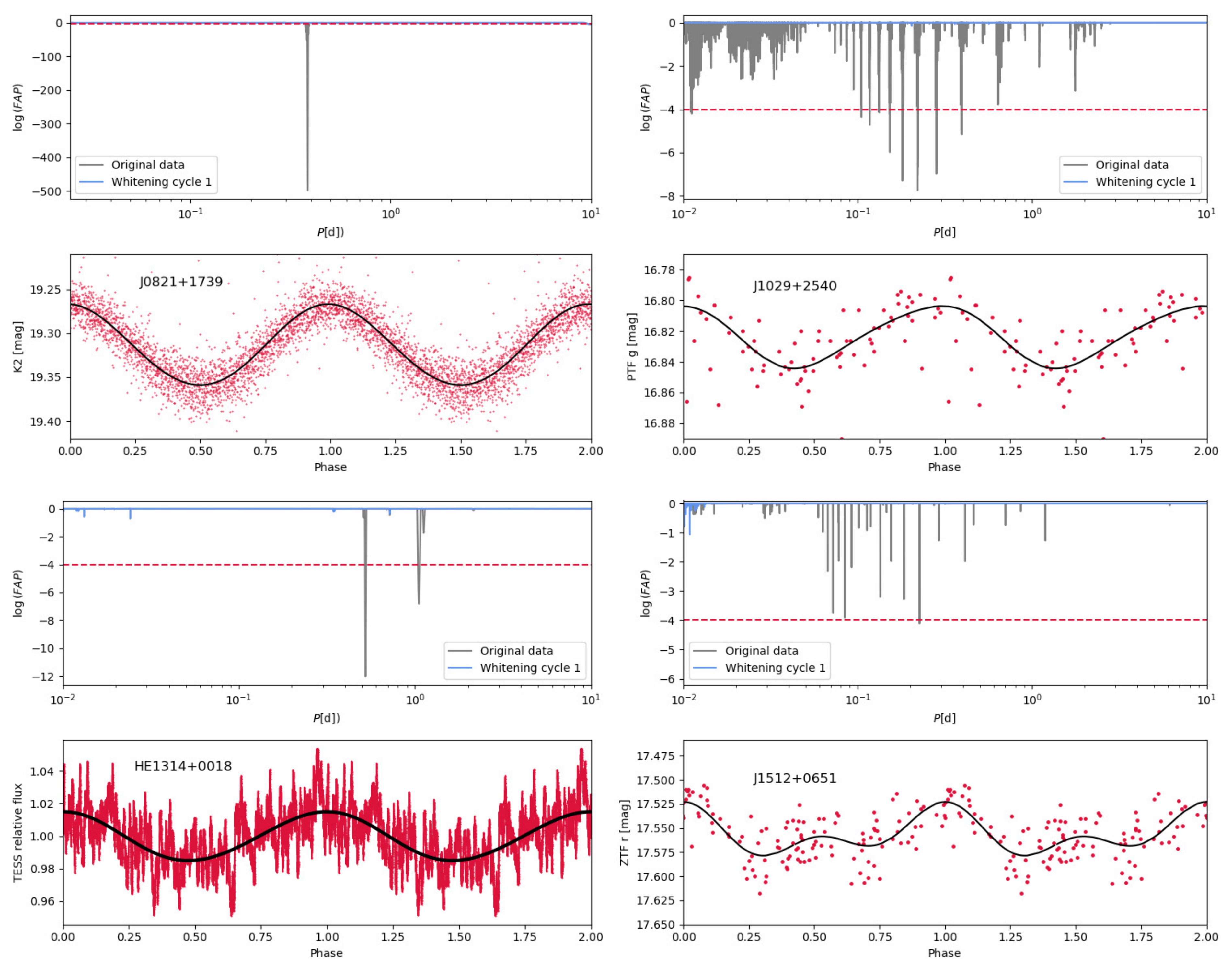}
\caption{Like \fg{fig:lp-1} for the He\,{\sc ii} line problem white dwarfs J0821+1739 and
  J1029+2540, HE\,1314+0018, and J1512+0651.}
\label{fig:lp-4}
\end{figure*}
 
\subsection{Notes on individual objects}
\label{sect:notes}
\subsubsection{UHE white dwarfs}
\label{sect:uhenotes}
 
\paragraph{J0032$+$1604} is a DO-type UHE white dwarf with the strongest
UHE features. It was observed within CSS and ATLAS. The periodograms
of all light curves show the strongest signal around 0.91\,d.
\cite{Heinze+2018} reports twice the period ($P=1.81619$\,d).
The amplitudes of the light curve variations range from 0.05\,mag
to 0.07\,mag, but are not found to differ significantly. In the first two rows
on the left side of \fg{fig:lp-1}, we show the periodogram and phase-folded light
curve from the ATLAS c-band, which predicts lowest FAP. The original
periodogram is shown in gray and the whitened periodogram is shown in
light-blue. No other significant signal is left after whitening the light curve
for the 0.91\,d periodicity. The black line on top of the phase-folded
light curve (red) is a fit of a harmonic series used to predict the
peak-to-peak amplitude.
\paragraph{WD0101$-$182} This bright ($G=15.74$\,mag) DOZ-type UHE white dwarf
was observed with TESS, CSS and ATLAS. The periodogram of
the TESS light curve shows the strongest peak around 2.32\,d. This period
is also confirmed by the CSS V band and ATLAS c band light curves,
respectively. The periodogram of the ATLAS o-band light curve predicts
the strongest peak at 1.747674\,d, but another significant
peak occurs at 2.31\,d, close to what is found in the
ATLAS c, CSS V, and TESS band. We also note, that the 2.32\,d periodicity
is already clearly visible in the unfolded TESS light curve and is also reported by
\cite{Heinze+2018}. The amplitudes of the ATLAS and CSS phase-folded light
curves are consistent.
\paragraph{J0146$+$3236} is the only object for which
rapid changes in the EWs of the UHE features were observed thus far.
\cite{Drake+2014} and \cite{Heinze+2018} already reported a photometric variability
of $P=0.484074$\,d (based on CSS data) and $P=0.48408$\,d (based
on ATLAS data), respectively, while \cite{Reindl+2019} reported half of that value.
We can confirm the period found by \cite{Reindl+2019}
based on ATLAS, ZTF, and TESS data. The periodogram of the
TESS light curve shows the strongest signal at $P=0.242037$\,d.
All other significant peaks turned out to be
(sub-)harmonics of this period (\fg{fig:lp-1}). The shapes of the phase-folded light curves
is roughly sinusoidal, with extended flat minima.
\paragraph{HS\,0158$+$2335} was observed with CSS, ATLAS, ZTF, and TESS.
In the TESS light curve, we detect the strongest signal around 0.45\,d.
No other significant period is left after the first whitening cycle.
In the periodograms calculated for the ATLAS o-band (96 data points) and ZTF
g-band (43 data points) no significant periodic signal can be detected. In all
other light curves we also find a significant period at $P\approx 0.45$\,d.
The period found by us is confirmed by \cite{Drake+2014} who reported
$P=0.449772$\,d based on CSS DR\,1 data. \cite{Heinze+2018}, on the other
hand, reports twice the period ($P=0.899571$\,d) found by us.
The shape of the phase-folded light curves clearly shows two maxima, with
the first one being at phase 0.0, the second one being at approximately phase
0.6, and the minimum is located around phase 0.3.
\paragraph{J0254$+$0058} was observed within CSS, ATLAS, ZTF, TESS, and is
the only object in our sample included in the SDSS stripe 82 survey.
\cite{Becker+2011}, \cite{Drake+2014}, and \cite{Heinze+2018} report a period of about
2.17\,d for this object, based on SDSS stripe 82 (u, g, and r band),
CSS V band, and ATLAS o- and c band light curves, respectively.
The periodograms of the light curves of all surveys mentioned above predict
the strongest periodic signal at around 1.09\,d. The amplitudes of the
phase-folded light curves are always around 0.3\,mag and do not differ
significantly amongst the different bands. The shapes of the phase-folded
light curves are -- just as for J0146$+$3236 -- roughly sinusoidal, with a
broad and flat minima (top row left in \fg{fig:lp-2}). After whitening the TESS
light curve for the 1.09\,d periodic signal and its
(sub-)harmonics, we find one more significant peak around 1.3\,d (marked with
a ``x'' in \fg{fig:lp-2}) just above our variability threshold
($\log(FAP)=-4.4<-4$). After the second whitening cycle no other significant
peak is left in the periodogram.
\paragraph{HE\,0504$-$2408} is one of the objects showing the strongest UHE
features, and one of the brightest ($G=15.77$\,mag) stars in our sample. It
was observed in the course of the CSS (69 data points) and the
SSS (182 data points). The SSS light curve indicates that the star underwent a
brightening of 0.4\,mag from $MJD=53599$ to $MJD=53755$ and remained at
$V\approx15.65$\,mag. Using only data obtained after $MJD=53755$ we find a
period of 0.684304\,d with an associated $\log(FAP)=-3.4$. The amplitude of
the phase-folded light curve is 0.08\,mag, and its shape is sinusoidal.
We classify this star as possibly variable.
\paragraph{HS\,0713$+$3958} is yet another example whose phase-folded light
curves show extended, flat minima (second row right in \fg{fig:lp-2}). The
periodogram of the TESS light curve shows the strongest periodic signal around
$P=0.78$\,d (first row right in \fg{fig:lp-2}). No
other significant signal is left in the periodogram after whitening the light
curve for this periodicity. The strongest periodic signals in the CSS, and ZTF
g- and r-band light curves are also detected around 0.78\,d.
In the ATLAS c- and o-band light curve the strongest periodic signals
are found at 1.379916\,d  and 0.304796\,d,
respectively. However, we also find in both periodograms periodic signals
around 0.78\,d above our FAP threshold.
\cite{Heinze+2018} report a period of $P=0.609618$\,d,
twice of what we found as the strongest signal in the ATLAS o-band.
We adopt the 0.781646\,d period from the TESS light curve.\\
Ground-based infrared photometry by \cite{Napiwotzki+1997} revealed a
nearby star to HS\,0713+3958. \cite{Werner+2018a}, who recorded an
optical spectrum with the Hubble Space Telescope (HST) of this late type
star, determined a spectral type of M5V and found that the spectroscopic
distances of both stars agree within the error limits.
Comparing the fluxes of the HST spectrum of the M5V star with the SDSS
spectrum, we find that the flux of the cool star only dominates beyond
10\,000\,\AA, i.e. beyond the TESS filter pass band. This implies that the
periodicity found in our light curve analysis most likely originates
from the hot white dwarf and not from the cool star. Another interesting
fact is, that companions of spectral type M5 or later may easily be
hidden in the optical due to the still high luminosity of the white dwarf.
We also note that Gaia clearly resolved the white dwarf and the M5 star
(we calculate a separation of $1.0396\pm0.0005$ arcsec), hence it is not
possible that the two stars form a close binary.
\paragraph{HS\,0727$+$6003} The periodogram of the TESS light curve shows
the strongest periodic signal around $P=0.22$\,d (penultimate row right
in \fg{fig:lp-2}). No other significant period is
found after the first whitening cycle. The $\approx0.22$\,d period is also
found in the CSS, ATLAS c- and o-band, and ZTF g- and r-band
light curves. Again, the minima of the phase-folded light curves are broad
and flat. The amplitudes are all around 0.13\,mag and do not differ
significantly amongst the different bands. \cite{Drake+2014} gives a period
of $P=0.28437$\,d, higher than what we find. \cite{Heinze+2018} reports
twice our period ($P=0.442823$\,d).
With BUSCA we could record almost two phases, and found that the
amplitudes of the $U_B$, $B_B$, and $R_B$ band light curves ($0.128\pm0.014$\,mag,
$0.131\pm0.008$\,mag, and $0.128\pm0.011$\,mag, respectively) agree with each other as
well as with the amplitudes from the light curves from the other surveys.
\paragraph{HS\,0742$+$6520} is - like HE\,0504$-$2408 - one of the objects with
the strongest UHE features and found to be not significantly variable. It was
observed only 121 times in the course of the CSS. The TESS light curve predicts
the strongest peak at 0.281989\,d with an associated $\log(FAP)=-1.7$. The
phase-folded light curve has an amplitude of 0.01\,mag only. Thus, this star
is likely not variable.
\paragraph{J0900$+$2343}
is a faint ($G=18.79$\,mag) DA-type UHE white dwarf.
Visual inspection of the K2 light curves processed by the EVEREST and K2SSF
pipeline indicates that the data still suffer from systematic errors. Thus,
we discard the K2 data of this object from our analysis. The star was also
observed within the CSS (469 data points) and ZTF (only 44 data points in both
the g- and r-band), however, no significant periodic signal can be detected
in those light curves. The non-detection of a variability in this object may
be a consequence of the faintness of the star.
\paragraph{J1059+4043} is half a magnitude brighter ($G=18.34$\,mag) than
J0900$+$2343. In the periodogram of the ZTF g and r  band light curves (about 230
data points each) we detect the strongest periods around $P=1.41$\,d. The
phase-folded light curves have an amplitude of 0.08\,mag and their shapes
are roughly sinusoidal (bottom row right in \fg{fig:lp-2} for the ZTF g band
data). In the periodogram of CSS V-band light curve (315 data points) no
significant period can be found.
\paragraph{J1215+1203}: This faint ($G=18.17$\,mag) DO-type UHE white dwarf
was observed in the course of the CSS, and ZTF. The periodograms of all
these light curves show the strongest periodic signal at $P\approx0.60$\,d.
The shape of the phase-folded light curve is roughly sinusoidal (top
row, left in \fg{fig:lp-3}).
\paragraph{J1257+4220} is a DA-type UHE white dwarf and was observed in the course
of the CSS, ZTF, and ATLAS. While in the CSS V-band and ATLAS o-band
no significant periodic signal can be detected, the ZTF light curves and
ATLAS c-band light curves indicate the strongest periodic
signal at $P\approx0.43$\,d.
\cite{Heinze+2018} classified J1257+4220 as sinusoidal variable with much
residual noise and, again, reports twice the period ($P=0.857925$\,d)
found by us.
\paragraph{J1510+6106} is a DO UHE white dwarf and two minute cadence light
curves are available from four TESS sectors. There are no blends with other
stars in the TESS aperture or a contamination by nearby bright stars (\fg{fig:tpf}).
In the periodogram of the combined TESS light curve
TESS light curve we find one significant peak at 5.187747\,d
($\log(FAP)=-5.1$), however, this signal is not found in any individual
sector light curve. This white dwarf was also observed more than 500 times
in both the ZTF g- and r-band. In those light curves no
significant periodic signals can be found. Thus, we remain skeptical about
the five day period from the combined TESS light curve, and classify
this star only as possibly variable.
\paragraph{HS\,2027+0651} is a DO UHE white dwarf that was observed within
the ZTF. The periodogram of the ZTF g-band light curve indicates
$P\approx0.29$\,d. The amplitude of the phase-folded light
curve is 0.06\,mag, and its minimum is again broad and flat
(bottom left panel in \fg{fig:lp-3}).\\
\paragraph{HS\,2115$-$1148} is a DAO-type UHE white dwarf with very weak
UHE lines. The periodogram of the ZTF r-band (\fg{fig:lp-4}) predicts the
strongest signal around 1.32\,d. The amplitude of
the phase-folded light curve is 0.04\,mag.
 
\subsubsection{White dwarfs showing only the He\,{\sc ii} line problem}
\label{sect:HeIIpnotes}
 
\begin{figure}[ht]
\centering
\includegraphics[width=\columnwidth]{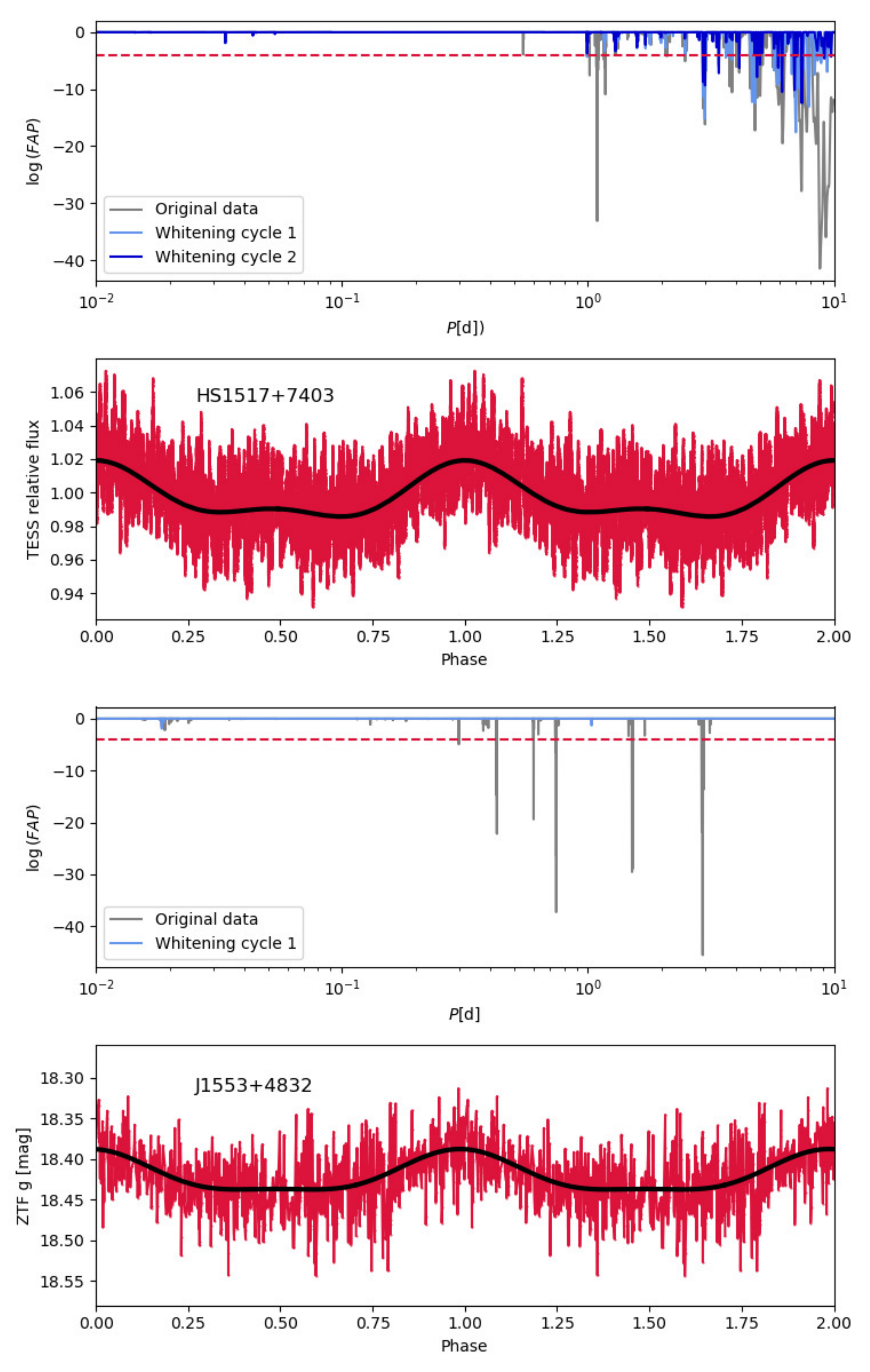}
\caption{Like \fg{fig:lp-1} for the He\,{\sc ii} line problem white dwarfs HS\,1517+7403 and J1553+4832.}
\label{fig:lp-5}
\end{figure}

\paragraph{J0821$+$1739} is the faintest object in our sample ($G=19.07$\,mag).
In the periodogram (top row left in \fg{fig:lp-4}) of the K2 light curve processed
by the EVEREST pipeline only one strong signal can be found at $P=0.384875$\,d.
This variability is already obvious from the (unfolded)
light curve. We note, that both the
amplitude and shape of the phase-folded K2 light curve must not be regarded as
reliable due to the long exposure time (5\% of the period). The
$\approx 0.38$\,d period is also confirmed by the K2 light curve processed by
the K2SFF pipeline, though, we obtain a higher $FAP$ for
the variability. Even though the target is quite faint, we also find the
$\approx 0.38$\,d period in the CSS and ZTF g-band light curves, however, in
the latter it is not significant ($\log(FAP)=-3.0<4$). The amplitude of the
phase-folded CSS light curve is 0.13\,mag.
\paragraph{J0827+5858} was observed 332 times in the course of the CSS
($V=17.46$\,mag), about 200 times in both the ZTF g- and r-band. We do
not find a significant periodic variability in any of those light curves.
\paragraph{J0947+1015} was observed 447 times in the course of the CSS
($V\approx18.07$\,mag), and 64/81 times the ZTF g/r band, respectively. The
periodogram of the CSS light curve indicates a period of 0.257938\,d
with an associated $\log(FAP)=-3.6$. The amplitude of the phase folded light
curve is 0.10\,mag. We classify this star as possibly variable.
\paragraph{J1029+2540}
In the periodogram of the ZTF g-band light curve we find the strongest periodic
signal in the ZTF g-band around $P=0.28$\,d (first row right in
\fg{fig:lp-4}). This period is confirmed by the  CSS V-band and ZTF r-band.
\paragraph{HE\,1314+0018} In the TESS data of this fairly bright ($G=16.05$\,mag)
star we find a significant period around $0.52$\,d. The
amplitude of the phase folded light curve is only 0.03\%. After the first
whitening cycle no other significant peak remains in the periodogram (penultimate row
left of \fg{fig:lp-4}). The star was also observed 368 times within the CSS,
however, in this data set no significant periodic signal can be found.
\paragraph{J1512+0651}
Was observed 103/119 times in the ZTF g/r band, and 365 times in the CSS
V-band. In the periodogram of the ZTF r band we find the strongest signal
at 0.226\,d. In the ZTF g and CSS V band we also find the 0.226
period, however, at FAPs below our threshold. The amplitude of the
phase-folded ZTF r band light curve is 0.06\,mag.
\paragraph{\object{HS\,1517+7403}}
In the periodograms of the ZTF g- and r-band light curves we find the strongest
signals around 1.09\,d, respectively. After the first whitening cycle, no
other significant signal remains. The star was also observed with TESS.
The periodogram of the TESS light curve predicts the strongest peak around
8.78\,d, however, another strong signal is detected at
1.09\,d confirming what is found from in the ZTF light
curves. Since in the ZTF periodograms we do not see a significant peak at around
8.78\,d, we adopt 1.09\,d as the photometric period of the star.
After whitening the TESS light curve for the 1.09\,d period (including it
harmonics and subharmonics), the signal at 8.78\,d disappears, however,
other significant signals around 7\,d, and 2\,d remain. Since those signals
are not detected in the ZTF periodograms, we conclude that they most likely
originate from the other star(s) inside the aperture mask, or the two orders of
magnitude brighter star right next to it (bottom row right in \fg{fig:tpf}).
\paragraph{J1553+4832}
This faint (G=18.65\,mag) object was observed about 1200 times in
course of the ZTF. In both, the periodograms of the ZTF g and r band, we
find the strongest signals around 2.93\,d. The amplitudes of the phase-folded light
curves in both bands is about 0.05\,mag. We note, that there are also
aliases at lower periods (e.g. at 1.52\,d and 0.74\,d), which have a
similar FAP (all of them are removed after the first whitening cycle).
Thus, it may be possible that the real photometric period is shorter. The star was
also observed 171 times within the CSS, however, in this light
curve no significant periodic signal can be found.
 
\section{Overall results}
\label{sect:results}
 
\subsection{Variability rates}
\label{sect:var}
We find that 12 out of the 16 UHE white dwarfs are significantly
photometrically variable, meaning their light curves exhibit periodic signals
with a $\log(FAP)\leq-4$. This leads to a variability rate of $75^{+8}_{-13}$\%.
Given the low-number statistics, the uncertainties were calculated assuming
a binomial distribution and indicate the 68\% confidence-level interval
(see e.g. \citealt{Burgasser+2003}). For two objects, HE\,0504$-$2408 and
HS\,0742$+$6520, we find periodic signals with associated
$\log(FAPs)\approx -3$. For J1510$+$6106 we do not trust the signal
around 5.19\,d discovered in the combined TESS light curve, since
it can neither be found in the ZTF g or r band light curve (about 500 data
points each), nor in the four individual TESS light curves.
Those objects we consider as possibly variable.
For the DA-type UHE white dwarf J0900$+$2343 no hint for a variability
could be found, which, however, might be a consequence of the faintness of the star
($G=18.79$\,mag).
For the white dwarfs that show only the \Ion{He}{2} line problem, we find
a similar variability rate of $75^{+9}_{-19}$\%, meaning that six out of the eight
\Ion{He}{2} line problem white dwarfs are significantly photometrically
variable. For J0827+5858 we cannot find a significant periodic signal, and
J0947+1015 we classify as possibly variable. The high photometric
variability rate amongst these stars suggests that the UHE and \Ion{He}{2}
line problem phenomena are linked to the variability.\\
But is the photometric variability indeed an
intrinsic characteristic of these stars alone, or not
rather something that is observed amongst \textit{all} very hot white dwarfs?
In order to test this, we obtained ZTF DR4 light curves of a
comparison sample and search for photometric variability in those light
curves as well. Our comparison sample consist of several very hot (\Teff$\geq 65\,000$\,K)
DO-type (55 in total, including 28 PG1159-type stars) white dwarfs from
\cite{Dreizlerwerner1996, DreizlerHeber1998, Huegelmeyer+2005, Huegelmeyer+2006,
  WernerHerwig2006, Reindl+2014c, Werner+2014} and \cite{Reindl+2018a}, as well as
very hot (\Teff$\geq 65\,000$\,K) DA-type (90 in total) white dwarfs from the
samples of \cite{Gianninas+2011} and \cite{Tremblay+2019}. We considered only
ZTF light curves which have at
least 50 data points (this value was found from our previous analysis
to be approximately needed to detect periodic signals in the ZTF data).
We find that amongst the H-deficient white dwarfs, only one of the 41 objects
with sufficient data points in the ZTF is significantly variable (variability
rate: $2^{+5}_{-1}$\%)\footnote{We note that the ZTF data are not suitable to
  detect pulsations. Otherwise a higher variability rate could be expected for
  very hot H-deficient white dwarfs, as many of them are GW Vir pulsators.}.
For the H-rich white dwarfs we find a higher variability rate of
$14^{+6}_{-3}$\% (59 stars had at least 50 data points and eight turned out to be significantly
variable). In Table~\ref{tab:periods3}, we list all of the normal white dwarfs which
we found to be variable based on the ZTF data, including the mean magnitudes,
derived periods, and amplitudes.
The variability rate of all normal white dwarfs together is then $9^{+4}_{-2}$\% and in
stark contrast to the combined variability rate of $67^{+8}_{-11}$\% based on ZTF data for
the UHE and \Ion{He}{2} line problem white dwarfs\footnote{Amongst the UHE and
  \Ion{He}{2} line problem white dwarfs 21 objects have at least 50 data points
in at least one ZTF band, and 14 of them turned out to be variable based on
the ZTF data.}. Thus, we conclude that periodic photometric variability is
indeed a characteristic of UHE and \Ion{He}{2} line problem white dwarfs.
 
\subsection{Light curve shapes}
\label{sect:shapes}
The shapes of the light curves are quite diverse. Some objects show near
perfect sinusoidal variations (e.g. HE\,1314+0018, J1029+2540),
while the light curves of seven objects in our sample (about one third
amongst the variable ones) show extended, flat minima (J0254+0058, J0146+3236, HS\,0713+3958,
HS\,0727+6003, HS\,2027+0651, J1553+4832, and HS\,1517+7403).
Particularly interesting are the light curves of HS\,0158+2335, that show
two uneven maxima. This might also be the case for J1512+0651 (shows only
the \Ion{He}{2} line problem), though, higher S/N light
curves would be needed to confirm this.
 
\subsection{Amplitudes}
 
\begin{figure*}[t]
\centering
\includegraphics[width=\textwidth]{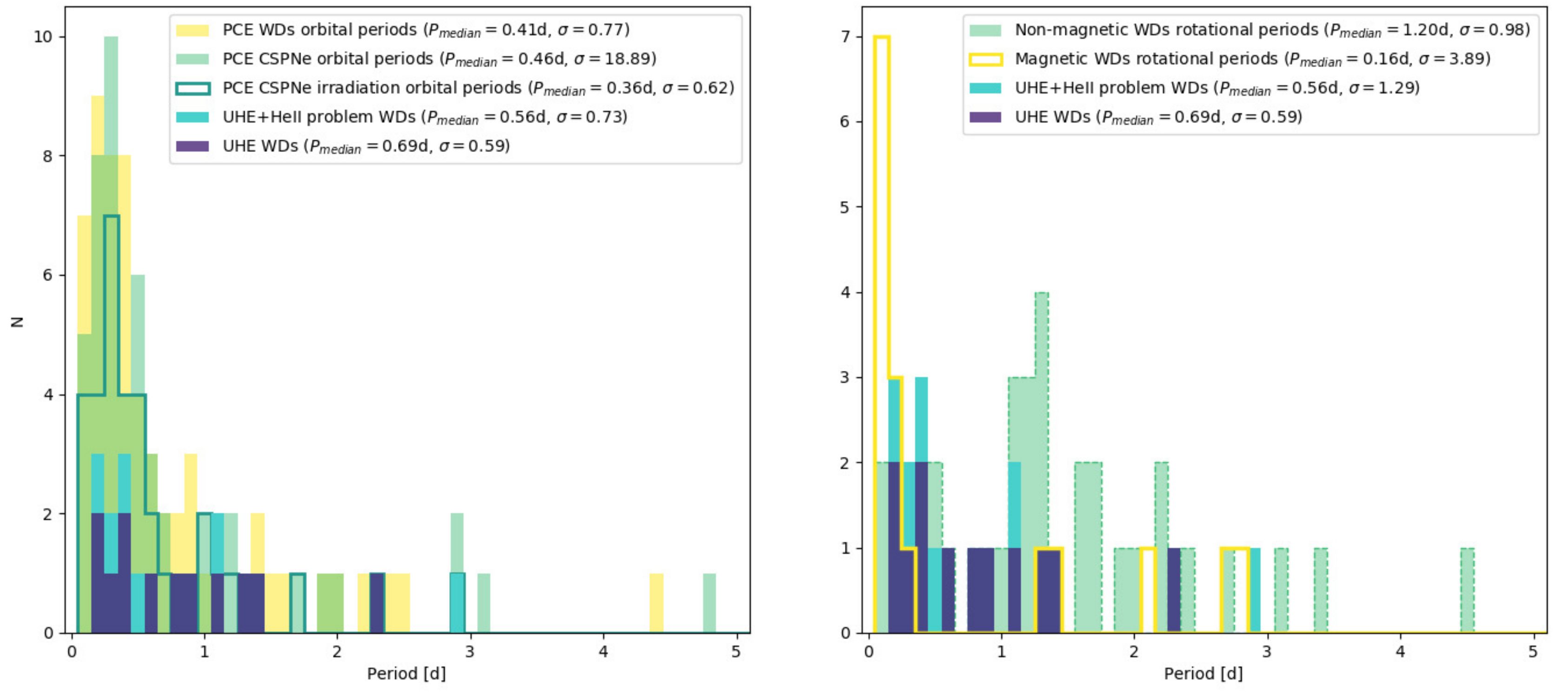}
\caption{Distribution of the photometric periods of the variable UHE and
  He\,II line problem white dwarfs (blue, in purple the period distribution
  of only the UHE white dwarfs is shown). In the left their period distribution
  is compared to the orbital period distribution of PCE CSPNe (light green,
  the bold teal line indicates the period distribution of binary CSPNe that
  show an reflection effect) and white dwarfs plus main sequence binaries
  (light yellow). In left panel a comparison with the rotational
  periods of normal white dwarfs (light green with dashed contours) and
  magnetic white dwarfs are shown (bold yellow lines). The median period
  and standard deviation of each sample is indicated.}
\label{fig:periods}
\end{figure*}
 
The amplitudes of the light curve variations range from a few tenth to
hundredth mag. For a given object, the amplitudes in the different bands
do not vary significantly. That means, we find that the difference in the
amplitudes as measured in the different bands, is smaller or equal than
the standard deviation of the difference between the observations and our
mathematical fit (black lines in \fg{fig:lp-1}-\fg{fig:lp-5}).
In particular, the SDSS stripe 82 light curves of
J0254+0058 (the only object in our sample with u till z band data), do
not indicate an increase of the amplitudes towards shorter or longer
wavelengths. Also in the BUSCA light curves of HS\,0727+6003 (only other
object with U-band light curve) we could not
observe any hint for a difference in the amplitudes.\\
We note that we do not trust the amplitudes of the TESS light curves. This is
because the TESS mission was designed for stars brighter than 15\,mag, and all
our targets are fainter than this. Second, the large pixel size implies that
an accurate background subtraction is very complicated, in particular in
crowded fields. The majority of the TESS light curves
predicts amplitudes that are larger than what is observed in the other bands.
For example, the amplitude of the phase folded TESS light curve of J0254+0058
is 0.54\,mag -- almost twice of what is observed in the other bands
($\approx 0.3$\,mag). If the large TESS amplitude would be actually real, then
we would expect to see similarly large amplitudes in the SDSS i- and z-band
as well, but which is not the case.
The faintness of our targets and large TESS pixel size of 21\,arcsec, which often
leads to contamination from neighboring stars, also results in a large scatter in the
TESS light curves. This in combination with the shorter duration of the
TESS light curves compared to those obtained from ground-based surveys like
ZTF (about one month compared to more than two years), explains the
larger uncertainties on the periods obtained from the TESS data.

\subsection{Periods}
 
The photometric periods of the UHE white dwarfs range from 0.22 to
2.32\,d, with a median of 0.69\,d and a standard deviation of 0.59\,d.
For the six photometrically variable white dwarfs showing only the \Ion{He}{2} line
problem, we find a very similar period range from 0.22 to 2.93\,d, with a
median of 0.45\,d and a standard deviation of 0.95\,d. Considering both classes
together we find a median of 0.56\,d with a standard deviation of 0.73\,d.\\
The observed periods are consistent with typical white dwarf rotational rates
\citep{Kawaler2004, Hermes+2017}, but could also indicate post-common
envelope (PCE) binaries \citep{Nebot+2011}. It is therefore worth comparing
the period distribution of those objects to the period distribution of our
sample in detail.\\
In \fg{fig:periods} we show in blue the combined period distribution of
the UHE white dwarfs and white dwarfs showing only the \Ion{He}{2} line
problem. In purple the period distribution of only the UHE white dwarfs is
shown. In the left panel we compare their period distribution to the orbital
period distribution of confirmed post-common envelope (PCE) binary central stars
of planetary nebulae (CSPNe, light green,
\citealt{Jones+Boffin2017, Boffin+Jones+2019})\footnote{\url{http://www.drdjones.net/bcspn/}} and
PCE white dwarf and main sequence binaries (light yellow) from the sample of \cite{Nebot+2011}.
The bold teal line indicates the period distribution of binary CSPNe that
show a reflection effect. In the right panel we show a comparison with the
rotational periods of pulsating white dwarfs (light green with dashed
contours, values taken from \citealt{Kawaler2004, Hermes+2017}) and apparently
single magnetic white dwarfs (bold yellow lines, values taken from \citealt{Ferrario+2015}).
We note that there are also a few longer period magnetic white dwarfs
\citep{Putney+Jordan1995, Bergeron1997, Schmidt1999, Kawka+Vennes2012}
and PCE binary central stars \citep{Miszalski+2018b, Miszalski+2018a, Brown+2019},
which we omit from \fg{fig:periods} for better visibility.
From this figure it already seems that the period distribution of our sample
resembles more the period distribution of PCE binaries than
the rotational period distribution of white dwarfs. The median rotational period
of non-magnetic white dwarfs is 1.20\,d, while the median period of our sample
is half of that.
The observed rotational periods of magnetic white dwarfs as determined from
polarimetry and photometry range from a few minutes, to hours, to days,
over decades to centuries. The short spin period ones show their peak near
0.1\,d, a period much shorter than what we observe for the UHE white dwarfs and
white dwarfs showing the \Ion{He}{2} line problem.
 
In order to test the statistical significance of this impression
we performed two-sample Kolmogorov-Smirnov tests. This test allows
to compare two samples and to check the equality of their one-dimensional
probability distributions without making specific distributional
assumptions. The statistical analysis is based on a D-value that represents
the maximum distance between the empirical cumulative
distribution function of the sample and the cumulative distribution
function of the reference distribution. Based on the D-value, we then calculate the
p-value, which is used to evaluate if the outcomes differ significantly.
It is a measure for the probability of obtaining test results at
least as extreme as the results actually observed, assuming that the null
hypothesis is correct. In our case the null hypothesis is that the
two samples which are compared follow the same distribution. A p-value
of one indicates a perfect agreement with the null hypothesis, while a
p-value approaching zero rejects the null hypothesis.
We performed these tests for the various samples mentioned above.
First, we find that the period distributions of both UHE white dwarfs,
and white dwarfs showing only the \Ion{He}{2} line problem agree with each
other ($p=1.00$). We also find that the period distribution of our sample
agrees with that of PCE white dwarf plus MS binaries ($p=0.42$) and
PCE CSPNe ($p=0.60$ for all binary CSPNe and $p=0.25$ for only the binary
CSPNe showing a reflection effect).
No agreement is found with the rotational period distribution of magnetic
($p=0.007$), and non-magnetic white dwarfs ($p=0.04$).\\
We should, however, keep in mind that the stars in our sample are in earlier
evolutionary stages compared to the white dwarfs with measured rotational
periods. According to \cite{Althaus+2009} the radius of a DO white dwarf with
typical mass of 0.6\,\Msol\ decreases from 0.017\,\Rsol\ to
0.013\,\Rsol\ while the star cools down from 80\,000\,K (typical \Teff for a
UHE white dwarf) to 20\,000\,K (the majority of magnetic
white dwarfs from \citealt{Ferrario+2015} are reported to have temperatures
below this value, as well as all of the non-magnetic white dwarfs from
\citealt{Kawaler2004, Hermes+2017}). If we assume conservation
of angular momentum, then the rotational period should
decrease approximately by a factor of 0.5.
Therefore, we repeated the statistical tests under the simplified assumption
that all of the objects in our sample will halve their periods
as they cool down. By that we find that there is no agreement with
the rotational period distribution of non-magnetic white dwarfs ($p=0.0001$), but
a statistically meaningful agreement with the rotational period distribution
of magnetic white dwarfs ($p=0.11$).
 
\section{Discussion}
\label{sect:discussion}
 
We found that both UHE and \Ion{He}{2} line problem white dwarfs
overlap in a narrow region in the Gaia HRD. As expected, they lie
on top of the white dwarf banana and are well separated from the
hot subdwarf stars, and are much bluer than similarly hot white dwarfs
with M dwarf companions.
On average, UHE white dwarfs are found to be slightly bluer and have slightly
brighter absolute G-band magnitudes than the white dwarfs showing only the
\Ion{He}{2} line problem. This might suggest that white dwarfs with UHE lines
could evolve into objects that show only the \Ion{He}{2} line problem.
However, better constraints on the temperatures of these stars as well as a
larger sample would be needed to investigate this possibility further.\\
Our light curves studies revealed that the majority of both the UHE
white dwarfs ($75^{+8}_{-13}$\%) and \Ion{He}{2} line problem white dwarfs
($75^{+9}_{-19}$\%) are photometrically variable. The fact that their
photometric period distributions agree with each other, and that their light curves
exhibit similar amplitudes and shapes, reinforced that both classes
are indeed related. What remains to be discussed is the cause of the
photometric variability and how it is linked to the occurrence of the
UHE features and \Ion{He}{2} line problem.\\
The photometric periods of all stars in our sample are well above the
theoretical upper limit of $10^4$\,s predicted for non-radial g-mode
pulsations that are frequently observed amongst PG\,1159 stars (most
of them having periods below 3000\,s,
\citealt{Quirion+2007, Corsico+2019, Corsico+2020}).
Thus, we see two possible scenarios that could instead account
for the photometric variability in our stars -- one is linked to close
binaries, and the other one related to magnetic fields.
 
\subsection{Binaries}
\label{sect:binaries}
 
Because of the very good agreement of the period distribution of
our stars with that of PCE systems, an obvious assumption is that
our stars are close binaries. Then a variety of physical processes
could lead to the observed periodic variability. We rule out that the
objects in our sample are (over-)contact binaries, since the light curves
of such systems have extended maxima and narrow (sometimes V-shaped)
photometric minima and also often two uneven minima
(e.g., \citealt{Miszalski2009, Drake+2014}). Also ellipsoidal deformation that
occurs in a detached system and in which one star is distorted due to the
gravity of its companion, can be ruled out as main source for the photometric
variability. This is because the amplitudes of the light curve variations
caused by ellipsoidal deformation in systems that contain a hot and compact
white dwarf and an extended companion are always much smaller than that from the
so called irradiation effect.
 
\begin{figure}
\includegraphics[width=\linewidth]{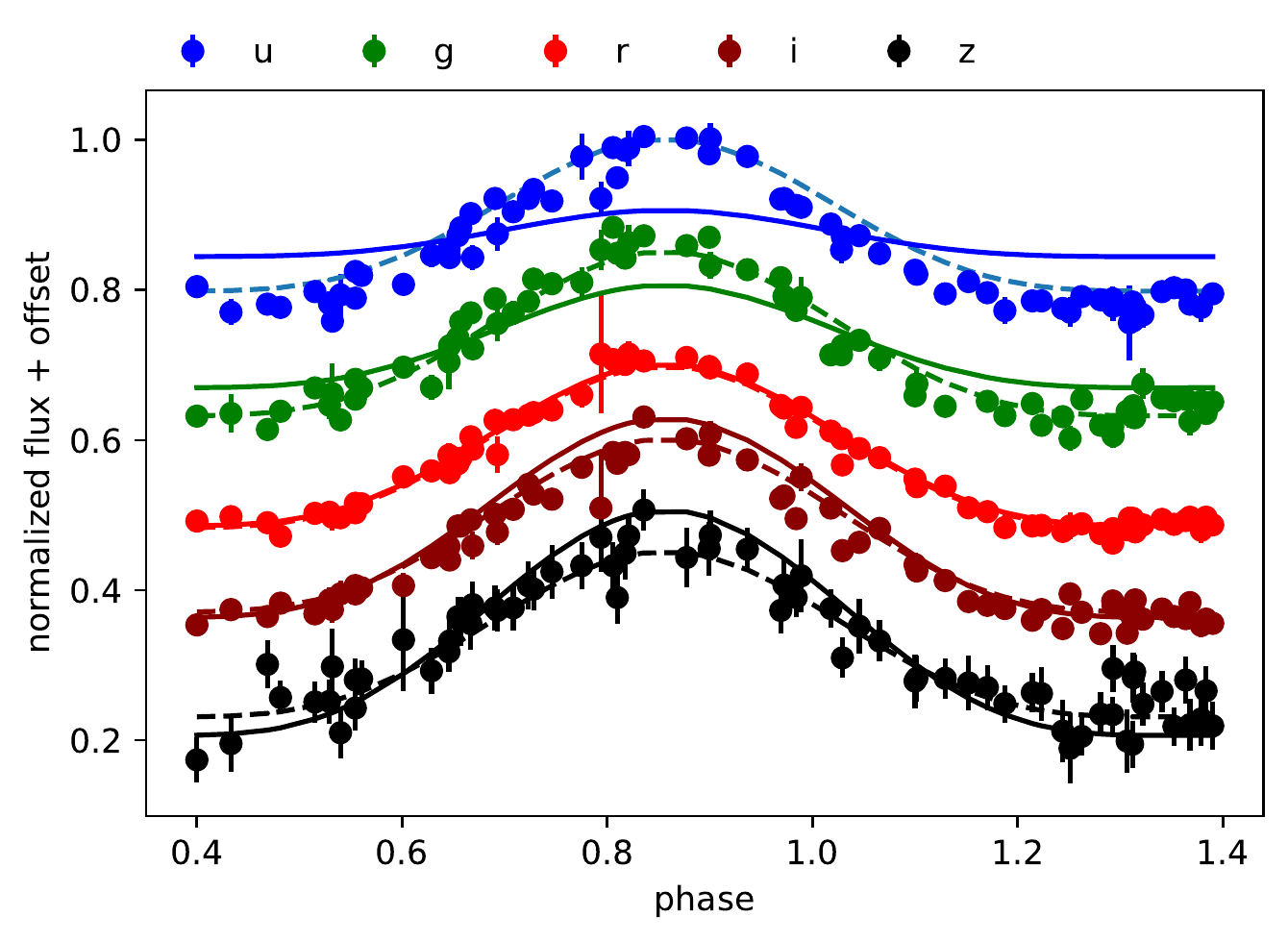}
\caption{SDSS-ugriz light curves of J0254+0058. The solid line shows the light curve models
using the parameters derived by fitting the SDSS-r light curve and a fixed
albedo of $A=1$ in all bands. The dashed lines give the light curve
model fit that allows for unphysical variations in the albedo of the companion.}
\label{lc_model}
\end{figure}
 
An irradiation or reflection effect, caused by the heated face (day-side)
of a cooler companion whose rotational period is synchronized
to the orbital period, appears as an attractive scenario.
Irradiation binaries display sinusoidal light curve variations,
however, when the system is seen under a
high inclination angle, the light curves have extended and flat
photometric minima, just what we find for seven objects in our
sample (\se{sect:shapes}). Well studied examples which exhibit
that latter kind of light curves are the hot subdwarf plus M-dwarf binary
\object{HS\,2333+3927} \citep{Heber+2004}, and the hot white dwarf
plus M-dwarf binaries HS\,1857+5144 \citep{Aungwerojwit+2007}
and NN\,Ser (which also shows eclipses, \citealt{Brinkworth+2006}).
The observed amplitudes can be as low as 0.01\,mag and reach up to
about 1\,mag \citep{Shimansky+2006, Brinkworth+2006}, covering the
observed amplitude range of our objects. However, we see serious
problems with the irradiation effect system scenario. First,
we would expect to find -- at least for some objects -- noticeable
differences in the amplitudes observed in the different bands.
For example in the very hot (\Teff$\geq49\,500$\,K) white dwarf plus
low mass main sequence star irradiation systems
\object{SDSSJ212531.92-010745.9}, and the central stars of
\object{Abell 63}, \object{V477 Lyr}, \object{ESO330-9},
\object{PN\,HaTr\,7}, the ratio of the R-band to V-band amplitude
ranges from 1.13 to 1.38, \citep{Shimansky2015, Afsar+Ibanoglu2008,
  Hillwig+2017}. \object{WD1136+667} and \object{NN\,Ser} even display
r-band to g-band amplitude ratios of 1.44 and 1.67, respectively
(this work, \citealt{Brinkworth+2006}). An even larger differences
in the amplitudes by a factor of almost 2 are expected when also
u-band photometry is available \citep{DeMarco+2008}. This should be
easily noticeable in the light curves of J0254+0058 and HS0727+6003.\\
In order to test this we calculated reflection effect models for the SDSS-ugriz
light curves of J0254+0058. We used the code \textsc{lcurve} \citep[for details, 
see  Appendix A in][]{Copperwheat+2010}, which was developed for white dwarfs
plus M-dwarf systems and has been used to fit detached or accreting white
dwarfs plus M-dwarf and hot subdwarf plus M-dwarf systems showing a
significant reflection effect \citep[see][for more details]{Parsons+2010,Schaffenroth+2020}.
For that we assumed \Teff$=80\,000$\,K for the white dwarf \citep{Huegelmeyer+2006} and
typical values for the masses and radii of white dwarfs plus M-dwarf systems
\citep[$q=0.21$, $R_1=0.02\,\rm R_\odot$, $R_2=0.15\rm\,R_\odot$, ][]{Parsons+2010}.
To find a first good model we fitted the SDSS-r light curve by letting the inclination $i$,
the temperature of the companion $T_2$, and the albedo of the companion vary. We found a
perfectly fitting model for an inclination of $i=86.8^\circ$ and a temperature of the companion
of $T_2=4500\,\rm K$. To see if this is also consistent in the other bands,
we fixed the stellar parameters of both stars and derived light curve models
for the other bands. We were only able to
fit the light curve, if the albedo of the companion was varied significantly
($A=0.6$ in SDSS-z to $A=3.5$ in SDSS-u, dashed line in Fig.\,\ref{lc_model}).
Such a large change in the albedo is unphysical, as the albedo gives the
percentage of the flux from the white dwarf that is used to heat up the
irradiated side of the companion.
If we assume a albedo $A=1$ the amplitude of the light curves varies
significantly from smaller in the blue to larger in the red, as shown in
Fig.\,\ref{lc_model}.\\
\begin{figure}
              \includegraphics[width=\columnwidth]{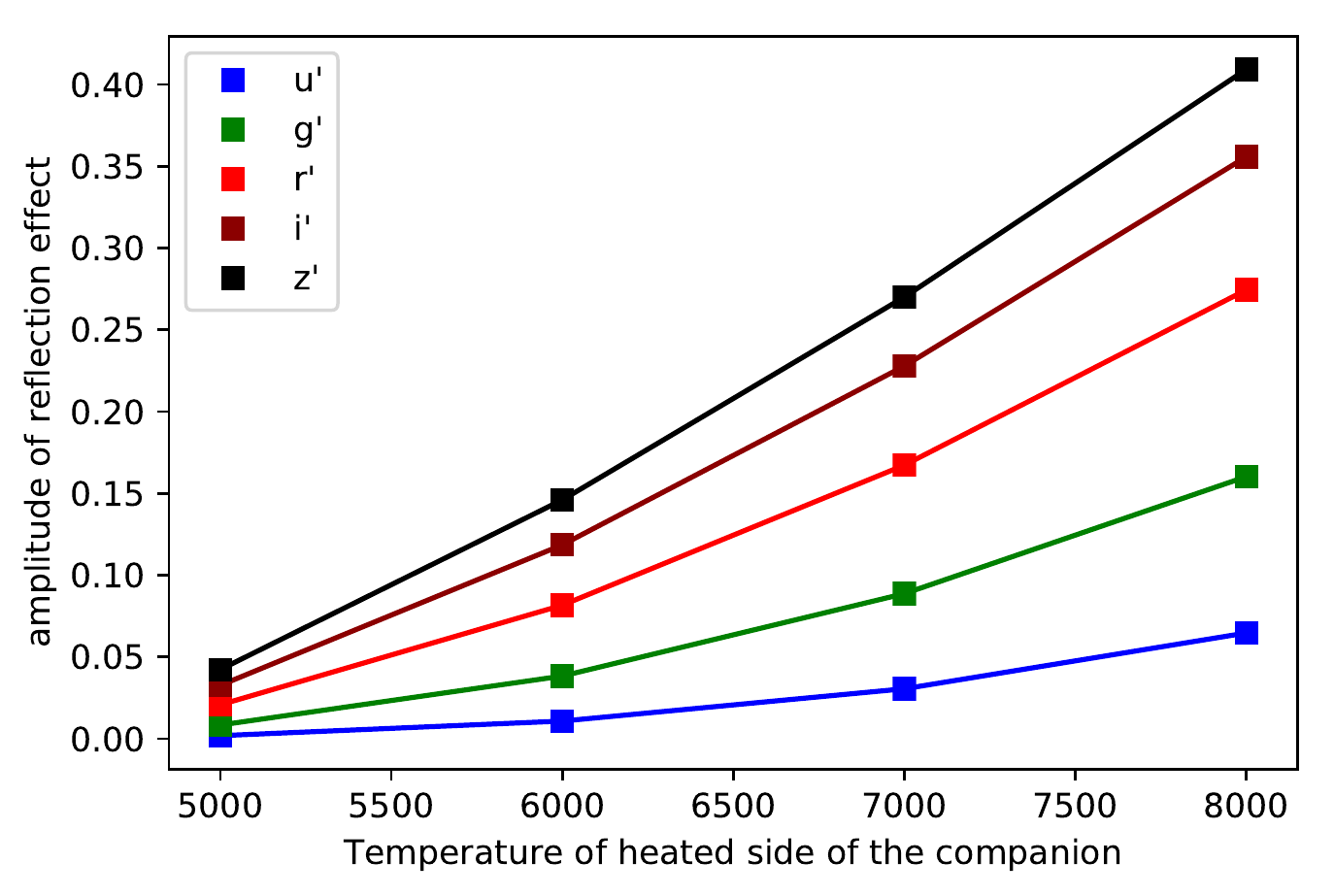}
              \caption{Expected amplitude for J0254+0058 of the reflection effect as a function of the temperature of the temperature of the heated side
of the companion. The amplitude was calculated by the difference in flux of a
white dwarf and a M-dwarf companion
with the parameters derived in the light curve fit in phase 0 and phase 0.5 using a black body approximation.}
              \label{amplitude}
\end{figure}
As explained before, this increase of the amplitude of the reflection effect from blue to red is
expected. The amplitude of the reflection effect is given by the difference in the flux between
phase 0, where the white dwarf and the maximum projected area of the cool side
of the companion is visible, and phase 0.5, where the white dwarf and the
maximum projected area of the heated side of the companion is visible. Depending on the temperature of
the white dwarf and the orbital separation of the system the companion is heated up to around
$10\,000-20\,000\,\rm K$. As the white dwarf has the maximum of the flux in the UV, the contribution of the
companion increases from blue to red.\\
To simulate this, we used the parameters that we derived in the light curve fit and used a black
body approximation to calculate the amplitude of the reflection effect as a
function of the temperature of the heated side of the companion. As the
period of the putative binary system is relatively long,
we calculated amplitudes up to 8000\,K for the heated side of the
companion. This is shown in Fig.\,\ref{amplitude}. A significant increase of
the amplitude from SDSS-u (5\%) to SDSS-z (40\%) is
predicted, which is not observed. From Fig. \ref{amplitude} it also becomes
clear, that the amplitude in the r band should be about twice of that in the g
band. However, also none of the ten other objects, which show significant periodic
variations in both ZTF bands, show an increased amplitude in the r band
compared to the g band.\\
The second drawback of the
reflection effect scenario is that none of our stars exhibits
spectral features of a cool secondary (\fg{fig:spectraUHE} and
\fg{fig:spectraHeIIP}). As mentioned before, a late-type M dwarf or a
brown dwarf may easily be outshined by the still luminous white dwarf,
thus the non-detection of an increased continuum flux in the optical or
lack of (molecular) absorption features from the companion cannot serve
as killer argument. However, to our
very best knowledge, without exception all PCE systems containing
a very hot (\Teff$\geq60\,000$\,K) white dwarf primary (and even those who
outshine their cool companions in the optical), exhibit emission
lines (e.g. the Balmer series or the CNO complex around 4650\,\AA)
arising from the highly irradiated hemisphere of secondary. These emission lines
are typically quite strong and can therefore also be detected in low
resolution (e.g. SDSS) spectra \citep{Nagel+2006, Nebot+2011}. It is also
well known that the emission lines appear and disappear over the orbital
cycle, reaching maximum strength at photometric maximum. Thus, it may be
possible, that when the systems is observed close to the photometric
minimum, that the emission lines are not detectable. But it is more than
unlikely that all spectra of the stars in our sample were taken at just
that phase.\\
For a reflection effect the amplitudes of the light curve
variations are expected to be correlated to the temperature of the day-side of
the irradiated companion. If we assume that all hypothetical close
companions to our stars have the same temperature, then the amplitudes should
correlate to $L/P^{2/3}$, where $L$ is the luminosity of the white dwarf and
P the orbital (photometric) period. This means that more luminous primaries
at shorter orbital periods are expected to cause a larger reflection effect
than less luminous primaries at longer periods. However, using $M_{G_0}$ as a proxy for
$L$, no correlation between $M_{G_0}/P^{2/3}$ and the mean amplitudes is
found (Pearson correlation coefficient: $r=-0.01$)\footnote{We only used
  objects with a relative uncertainty for the parallax smaller}
  than 20\% to check for this correlation.\footnote{The inclination angle
  of the system also has an impact on the amplitudes, which would cause an
  additional scatter. However, it is unlikely the inclinations are distributed
  in such a way that the correlation of the amplitude to $M_{G_0}/P^{2/3}$ just vanishes.}
This serves as a third argument against our stars being reflection effect
binaries.\\
Finally, we would like to note, that if the variability in all our objects
would be indeed caused by close companions, it would imply an
exceptionally high compact binary fraction amongst H-deficient stars
of 30\%\footnote{30\% of all DO-type white dwarfs hotter than 65\,000\,K
  show UHE lines or only the \Ion{He}{2} line problem. If we exclude those
that classify as PG\,1159 stars ($C/He>0.02$, number fraction) from the group
of normal DO-type white dwarfs a percentage of 47\% is obtained.}.
Amongst the immediate precursors of DO-type white dwarfs, only one O(He)
star and one luminous PG\,1159 star\footnote{Only ten O(He) stars and 16 PG\,1159
  pre-white dwarfs (\logg$<7.0$) are known.} are known to be
radial velocity variable \citep{Reindl+2016}. Another O(He)-type star, the
central star of Pa\,5 shows a photometric variability of 1.12\,d, which
however, might also be attributed to spots on its surface
\citep{DeMarco+2015}. Although no systematic search for close binaries
amongst these stars has been conducted yet, this would lead us to an estimated
close binary fraction of 11.5\% amongst H-deficient pre-white dwarfs, i.e.
a factor of 2.6 below what would be needed to explain the variability in our
stars via close binaries.
 
\subsection{Magnetic fields}
\label{sect:magneticfields}
 
The fraction of the hottest white dwarfs that show UHE lines or the
\Ion{He}{2} line problem (about 10\%) matches the fraction of magnetic
white dwarfs (2-20\% are reported, \citealt{Liebert+2003, Giammichele+2012,
  Sion+2014, Kepler+2013, Kepler+2015}). In addition, we found that the
period distribution of our stars agrees with that of magnetic white dwarfs
if we assume they will spin-up as a consequence of further contraction.
Proposing UHE white dwarfs are magnetic, \cite{Reindl+2019} suggested that
optically bright spots on the magnetic poles and/or geometrical effects of
a circumstellar magnetosphere could be responsible for the photometric
variability in J0146+3236.\\
Spots on hot white dwarfs are expected to be caused by the accumulation of metals
around the magnetic poles \citep{Hermes+2017c}. This is also the case for
chemically peculiar stars, where the magnetic field produces large-scale
chemical abundance inhomogeneities causing periodic modulations of spectral
line profiles and light curves \citep{Oksala+2015, Prvak+2015, Prvak+2020,
  Krticka+2018, Krticka+2020b}. This is understood as
a result from the interaction of the magnetic field with photospheric atoms
diffusing under the competitive effects of gravity and radiative levitation
\citep{Alecian2017}. If the radiative and gravitational forces are of similar
orders of magnitude, these structures are able to form and subsist
\citep{Wade2017}. In fact, it was found by \cite{Reindl+2014c}, that the
DO-type UHE and \Ion{He}{2} line problem white dwarfs are located at this
very region in the \Teff\ $-$ \logg\ diagram, where also the wind limit as
predicted by \cite{UnglaubBues2000} occurs. This further supports that
gravitational settling and radiation-driven mass loss hold balance in our
stars, and that, thus, long-lived spots can be expected.\\
\begin{figure}
              \includegraphics[width=\columnwidth]{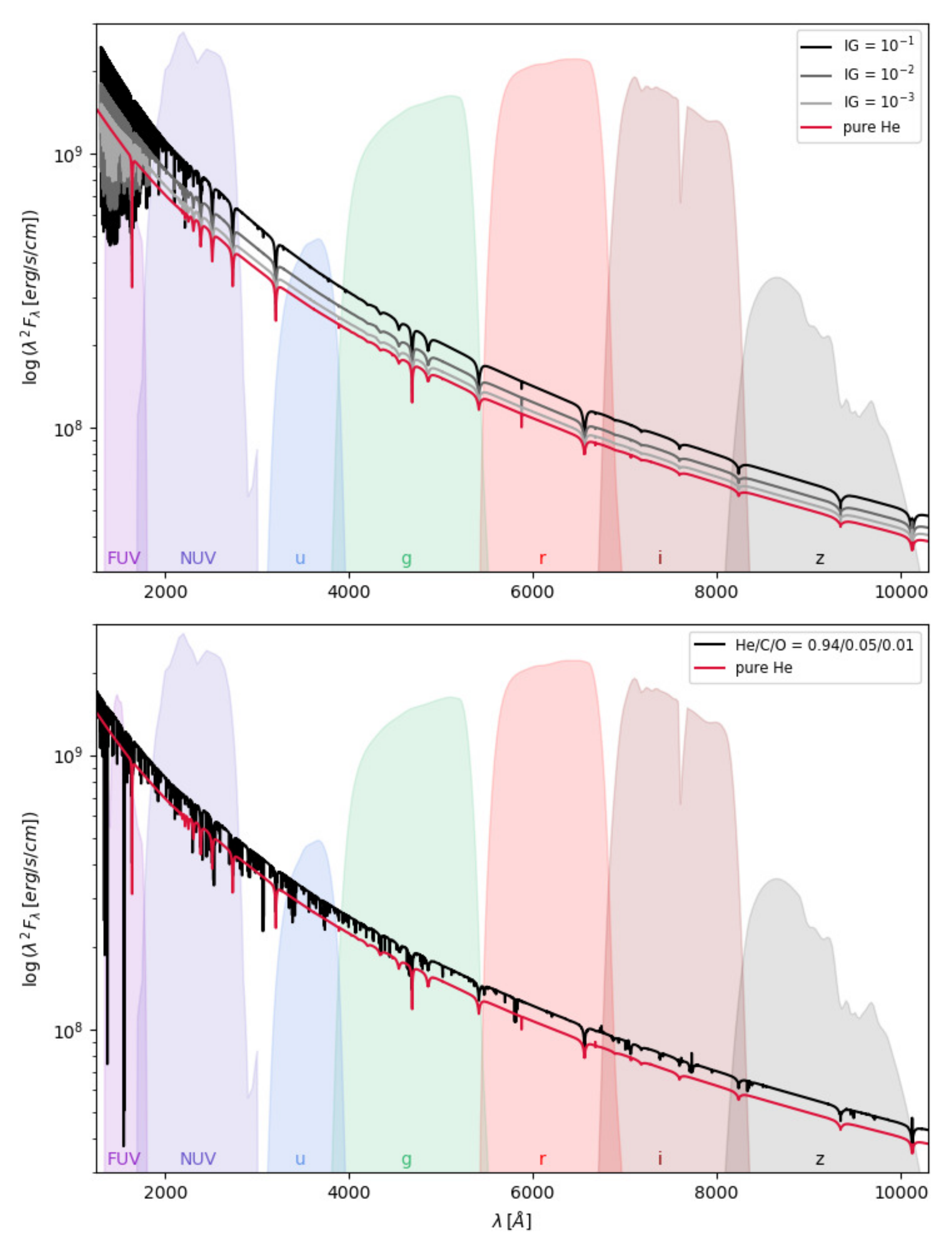}
              \caption{The differences in the fluxes of models with different
          metal contents and a model containing only He (red). The upper panel
          in shows fluxes for different abundances of the iron-group elements,
          and the lower panel shows a model that contains opacities of He, C,
          and O. The filter response functions of the Galex FUV, and NUV, as
          well as the SDSS u, g, r, i, and z bands are indicated.}
              \label{fig:fluxes}
\end{figure}
 
\cite{Reindl+2019} showed that the light curve of J0146+3236 can be modeled
assuming two uneven spots whose brightness is slightly over 125\% relative
to the rest of the stellar surface.
In order to get an idea about the metal enhancement needed to achieve
such an increase in brightness, we calculated test models with \textsc{TMAP}.
In the model atmosphere calculations, we assumed \Teff$=80\,000$\,K,
\loggw{8.0}, and included opacities of He and the iron-group elements
(Ca, Sc, Ti, V, Cr, Mn, Fe, Co, and Ni), of which Fe was found to be the
most abundant trace element in UHE white dwarfs \citep{Werner+2018b}.
Iron-group elements were combined in a generic model atom, using a
statistical approach, employing seven superlevels per ion linked by
superlines, together with an opacity sampling method
\citep{Anderson1989, rauchdeetjen2003}. Ionization stages
{\sc iv}-{\sc vii} augmented by a single ground-level stage {\sc viii} were
considered and we assumed solar abundance ratios. The models were calculated
for a metallicity of $10^{-3}$, $10^{-2}$, and $10^{-1}$ (mass fractions).
In addition, we calculated a model including besides He also opacities of C,
and O at typical abundance values of low-luminosity PG\,1159 stars (mass
fractions of $5\times10^{-2}$, and $1\times10^{-2}$, respectively). For the calculations we considered
ionization stages {\sc iii}-{\sc v} and {\sc iii}-{\sc vii} for C and O,
respectively, and a total of 404 non-LTE levels. Finally, also a pure He
model was computed. After that, the model fluxes were convolved with
filter response functions of the Galex FUV, and NUV, as well as the
SDSS u, g, r, i, and z bands to calculate synthetic magnitudes.\\
In \fg{fig:fluxes} the various synthetic spectra are shown, and the filter
response functions are indicated. The differences in the resulting magnitudes
relatively to our pure He model are listed in Table~\ref{tab:spots}.
We find that with an increasing abundance of the iron-group elements, the
continuum flux becomes steeper towards the UV. Most of the bound-bound
transitions are located at FUV wavelengths at this effective temperature,
which in turn causes a flattening of total flux in the FUV
band (upper panel in \fg{fig:fluxes}). Comparing our pure He model to our
model that contains also C and O, we find that the continuum flux also
increases from the near IR until FUV (hence also producing optically bright spots).
However, since many strong bound-bound transitions of C and O are located in
the optical (especially in the SDSS g band, lower panel in \fg{fig:fluxes}), the behavior of the
amplitude differences varies quite a bit from our models with iron-group
elements. This has been shown for \object{a Cen} by \cite{Krticka+2020a},
where for example an enhancement in He, Si, or Fe not only
predicts a different amplitude, respectively, but also the maxima of the
light curve variations are found to occur at different \textit{phases}.\\
We also note, that since spots cover only a part of the stellar surface,
the amplitudes listed in Table~\ref{tab:spots} can be seen merely as an upper
limit of what could be expected observationally from the metal enhancement
in the spot. Yet, it demonstrates that chemical spots could indeed
explain the relatively large amplitude variations we see in our stars.
The only drawback is, that for all metals considered here, the predicted
amplitude in the u band is always significantly larger than in the redder
bands. This is not observed for the two stars in our sample for which we have
u band light curves. However, only time-resolved UV spectroscopy combined
with detailed light curve modeling will be able to shed light on which
enhancement of elements could be responsible for the observed light
curve variability and if chemical spots are indeed the source of the
variability.\\
\begin{table}
\centering
\caption{Predicted differences in the resulting magnitudes from synthetic
  spectra containing metals relative to a model containing only He.
  The different photometric bands and metal abundances adopted in the
  calculations are listed.}
\label{tab:spots}
\begin{tabular}{c c c c c c}
\hline\hline
\noalign{\smallskip}
Model & IG     & IG     & IG     & C, O \\
      & $10^{-3}$ & $10^{-2}$ & $10^{-1}$ & $5\times10^{-2}$\\
      &        &        &        & $1\times10^{-2}$\\        
\noalign{\smallskip}
Band & $\Delta m$ & $\Delta m$ & $\Delta m$ & $\Delta m$ \\
& \,[mag] & \,[mag]   & \,[mag] & \,[mag]\\
\hline
\noalign{\smallskip}
FUV & 0.096 & 0.234 & 0.462 & 0.148 \\
NUV & 0.239 & 0.381 & 0.624 & 0.182 \\
u   & 0.223 & 0.339 & 0.535 & 0.176 \\
g   & 0.037 & 0.133 & 0.296 & 0.089 \\
r   & 0.062 & 0.146 & 0.289 & 0.160 \\
i   & 0.040 & 0.117 & 0.247 & 0.103 \\
z   & 0.061 & 0.133 & 0.249 & 0.168 \\
\hline
\noalign{\smallskip}
\end{tabular}
\end{table}
 
Besides a chemically inhomogeneous photosphere, stellar magnetism can create
another source of photometric variability.
\cite{Munoz+2020} recently hypothesize, that the photometric variability
observed in magnetic O-type stars is a consequence of electron scattering
in the obliquely rotating magnetosphere, which periodically occults the
stellar disk. They presented theoretical light curves for various
inclinations, $i$, and magnetic obliquity angles, $\beta$, mass-feeding rates,
magnetic field strengths, terminal wind velocities, and smoothing lengths.
Increasing the latter four parameters, they find that the amplitude of the
light curves variations should increase. For low inclination and obliquity
angles, they find roughly sinusoidal light curve variations. When
$i+\beta>90$\textdegree, the magnetic equator crosses the observer's
line-of-sight twice per rotation cycle and a second maximum in the light
curve shows up. Interestingly, for intermediate inclination and obliquity angles
(e.g. $i=\beta=50$\textdegree) their models predict a relatively long, and
almost flat photometric minimum, just what we observe for seven of our stars.
One of these stars is J0146+3236 for which \cite{Reindl+2019} already
suggested $i\approx\beta\approx45$\textdegree. Also, the models of
\cite{Munoz+2020} predict that the photometric minimum should occur, when
the circumstellar magnetosphere is seen edge-on, i.e. when the column density of
the magnetospheric material occulting the stellar disc is highest.
The magnetospheric occultation model, might even be able to explain the
extraordinary light curve of HS0158+2335, which exhibits two uneven maxima.
Its light curve resembles the one of \object{LMCe136-1}, which could be
reproduced by \cite{Munoz+2020} assuming a dipolar offset model.

\section{Conclusions}
\label{sect:conclusions}
 
Our work revealed exceptionally high photometric variability rates amongst
both UHE white dwarfs and white dwarfs that show only the \Ion{He}{2} line
problem, marking them as a new class of variable stars.
We found further evidence that both classes are indeed related,
as concluded from their overlap in the Gaia HRD, similar
photometric variability rates, light curves shapes and amplitudes, as well as
period distributions.
While an irradiation effect could explain their observed period distribution,
and the shapes of their light curves, we believe that this scenario is
unlikely. This is because we do not detect increasing amplitudes towards
longer wavelengths in any object, nor do we see emission lines arising from
the strongly irradiated side of a hypothetical close binary.
Instead, we hold on to the suggestion of \cite{Reindl+2019} that the
variability is caused by magnetic spots and/or the co-rotating, circumstellar
material.
 
Further investigations are needed for a profound understanding of
these special objects. A systematic search for radial velocity variations,
as well as an IR excess in combination with detailed light curve modeling
will help to decide if the close binary scenario can really be ruled out.
On the other hand, the spots/magnetosphere scenario can be checked with
spectro-polarimetric observations and time-resolved UV (that is where
photospheric metals can be detected) spectroscopy, which in turn
could reveal the magnetic field strengths and chemical spots, respectively.
Last but not least, the discovery that the majority of the UHE and
\Ion{He}{2} line problem white dwarfs are photometrically variable, provides
an important observational constraint to detect more of these systems.

\begin{acknowledgements}
  We thank Jiri Krti\v{c}ka, Thomas Kupfer, and JJ Hermes for
  helpful comments. We thank Stefan Dreizler
      for providing us with the TWIN spectrum of HS1517+7403.
      VS is supported by the \emph{Deut\-sche
      For\-schungs\-ge\-mein\-schaft, DFG\/} through grant GE 2506/9-1.
      IP acknowledges support from the UK's Science and Technology
      Facilities Council (STFC), grant ST/T000406/1.
      IP was partially supported by the \emph{Deut\-sche
      For\-schungs\-ge\-mein\-schaft, DFG\/} through grant GE2506/12-1.
      The BUSCA observing run was made possible with support from STFC grant ST/T001380/1.
      Some of the data presented in this paper were obtained
      from the Mikulski Archive for Space Telescopes (MAST). This research
      has made use of NASA's Astrophysics Data System and the SIMBAD
      database, operated at CDS, Strasbourg, France.
      Based on observations collected at the German-Spanish Astronomical
      Center, Calar Alto, jointly operated by the Max-Planck-Institut
      f\"ur Astronomie Heidelberg and the Instituto de Astrof\'{i}sica de
      Andaluc\'{i}a (CSIC).
      Based on data obtained from the ESO Science Archive Facility
      under request number nreindl/584030.
      The TMAD (\url{http://astro.uni-tuebingen.de/~ TMAD})
      and TIRO tool (\url{http://astro.uni-tuebingen.de/~TIRO}) used for
      this paper was constructed as part of the activities of the German
      Astrophysical Virtual Observatory.
      This work has made use of data from the European Space Agency (ESA) mission
      {\it Gaia} (\url{https://www.cosmos.esa.int/gaia}), processed by the {\it Gaia}
      Data Processing and Analysis Consortium (DPAC,
      \url{https://www.cosmos.esa.int/web/gaia/dpac/consortium}). Funding for the DPAC
      has been provided by national institutions, in particular the institutions
      participating in the {\it Gaia} Multilateral Agreement.
      The CSS survey is funded by the National Aeronautics and Space
      Administration under Grant No. NNG05GF22G issued through the Science
      Mission Directorate Near-Earth Objects Observations Program.  The CRTS
      survey is supported by the U.S.~National Science Foundation under
      grants AST-0909182 and AST-1313422.
      Based on observations obtained with the Samuel Oschin 48-inch Telescope
      at the Palomar Observatory as part of the Zwicky Transient Facility
      project. ZTF is supported by the National Science Foundation under Grant
      No. AST-1440341 and a collaboration including Caltech, IPAC, the
      Weizmann Institute for Science, the Oskar Klein Center at Stockholm
      University, the University of Maryland, the University of Washington,
      Deutsches Elektronen-Synchrotron and Humboldt University, Los Alamos
      National Laboratories, the TANGO Consortium of Taiwan, the University of
      Wisconsin at Milwaukee, and Lawrence Berkeley National Laboratories.
      Operations are conducted by COO, IPAC, and UW. 
      This work includes data from the Asteroid Terrestrial-impact
      Last Alert System (ATLAS) project. ATLAS is primarily funded to search
      for near earth asteroids through NASA grants NN12AR55G, 80NSSC18K0284,
      and 80NSSC18K1575; byproducts of the NEO search include images and
      catalogs from the survey area. The ATLAS science products have been made
      possible through the contributions of the University of Hawaii Institute
      for Astronomy, the Queen's University Belfast, the Space Telescope
      Science Institute, and the South African Astronomical Observatory.
      This paper includes data collected by the TESS mission. Funding for the
      TESS mission is provided by the NASA Explorer Program.
      This work made use of \texttt{tpfplotter} by J. Lillo-Box (publicly
      available in \url{www.github.com/jlillo/tpfplotter}), which also made
      use of the python packages \texttt{astropy}, \texttt{lightkurve},
      \texttt{matplotlib} and \texttt{numpy}.
      IRAF is distributed by the National Optical Astronomy Observatory,
      which is operated by the Association of Universities for Research in
      Astronomy (AURA) under a cooperative agreement with the National
      Science Foundation.
      Funding for the Sloan Digital Sky Survey IV has been provided by the
      Alfred P. Sloan Foundation, the U.S. Department of Energy Office of
      Science, and the Participating Institutions.
      SDSS-IV acknowledges support and resources from the Center for High
      Performance Computing at the University of Utah. The SDSS
      website is \url{www.sdss.org}.
      SDSS-IV is managed by the Astrophysical Research Consortium
      for the Participating Institutions of the SDSS Collaboration including
      the Brazilian Participation Group, the Carnegie Institution for Science,
      Carnegie Mellon University, Center for Astrophysics | Harvard \&
      Smithsonian, the Chilean Participation Group, the French Participation Group,
      Instituto de Astrof\'isica de Canarias, The Johns Hopkins University, Kavli Institute for the
      Physics and Mathematics of the Universe (IPMU) / University of Tokyo, the Korean Participation Group,
      Lawrence Berkeley National Laboratory, Leibniz Institut f\"ur Astrophysik
      Potsdam (AIP),  Max-Planck-Institut f\"ur Astronomie (MPIA Heidelberg),
      Max-Planck-Institut f\"ur Astrophysik (MPA Garching), Max-Planck-Institut f\"ur
      Extraterrestrische Physik (MPE), National Astronomical Observatories of
      China, New Mexico State University, New York University, University of
      Notre Dame, Observat\'ario Nacional / MCTI, The Ohio State University, Pennsylvania State
      University, Shanghai Astronomical Observatory, United Kingdom Participation Group,
      Universidad Nacional Aut\'onoma de M\'exico, University of Arizona, University of Colorado Boulder,
      University of Oxford, University of Portsmouth, University of Utah, University of Virginia, University
      of Washington, University of Wisconsin, Vanderbilt University,
      and Yale University.
 
\end{acknowledgements}
 
\bibliographystyle{aa}
\bibliography{UHE}
 
\begin{appendix}
 
\section{Tables}
 
\begin{table*}
\caption{Periods, mean magnitudes and amplitudes as derived from various light cures for
  all periodically variable UHE white dwarfs.}
\label{tab:periods}
\begin{tabular}{l l r r c c l}
\hline\hline
\noalign{\smallskip}
Name & Band & Datapoints & Magnitude & $P$ & Amplitude & Comment \\
     &      &            &  [mag]    & [d] & [mag]     &         \\  
\hline
\noalign{\smallskip}
J0032+1604&    CSS        &           335        &           15.71    &              $0.907846\pm0.000090$&        0.05 & DOZ UHE\\
              &           ATLAS-c              &           119        &           15.73    &              $0.907871\pm0.000080$&        0.07 & \\
\noalign{\smallskip}
WD0101$-$182&           CSS        &           154        &           15.83    &       $2.323148\pm0.000107$&        0.18 & DOZ UHE \\
        &   ATLAS-c              &           166        &           15.72    &              $2.323235\pm0.000140$&        0.19 & \\
              &           ATLAS-o             &           170        &           16.17    &              $2.323285\pm0.000219$&        0.19 & \\
        &   TESS      &           15985   &                         &              $2.322138\pm0.001939$&        &           \\
\noalign{\smallskip}
J0146+3236&    CSS        &           333        & 15.59 & $0.242037\pm0.000002$ & 0.18 &DO UHE \\
              &           ATLAS-c              &           123        & 15.54 & $0.242035\pm0.000003$ & 0.17 & \\
              &           ATLAS-o             &           124        & 16.01 & $0.242036\pm0.000003$ & 0.16 & \\
               & ZTF-g  &       222     & 15.37 & $0.242038\pm0.000001$ & 0.18 &  \\
               & ZTF-r  &       279     & 15.91 & $0.242037\pm0.000001$ & 0.16 &  \\
               & ZTF-i  &        22     & 16.37 & $0.242057\pm0.000029$ & 0.17 &  \\
        &   TESS      &    12936          &           & $0.242037\pm0.000010$ &        &              \\
\noalign{\smallskip}
HS\,0158+2335&            CSS        & 332    & 16.83 & $0.449773\pm0.000005$ & 0.17 & DO UHE \\
              &           ATLAS-c              & 105    & 16.91 & $0.449817\pm0.000035$ & 0.22 &  \\
              & ZTF-g   & 206   & 16.79 & $0.449776\pm0.000004$ & 0.24 &  \\
              & ZTF-r   & 236   & 17.23 & $0.449783\pm0.000005$ & 0.21 &  \\
        &   TESS      & 12891             &           & $0.449767\pm0.000471$ & & \\
\noalign{\smallskip}
J0254+0058&    CSS        &    336  & 17.39 & $1.087163\pm0.000021$ & 0.26 & DO UHE                            \\
              &           ATLAS-c              &    114  & 17.39 & $1.087221\pm0.000074$ & 0.28 &  \\
              & ZTF-g   &    250  & 17.25 & $1.087160\pm0.000015$ & 0.30 &  \\
              & ZTF-r   &    263  & 17.73 & $1.087168\pm0.000027$ & 0.30 &  \\
              &           SDSS-u &    72     & 16.73 & $1.087148\pm0.000006$ & 0.26 &  \\
              &           SDSS-g  &    72     & 17.15 & $1.087145\pm0.000002$ & 0.27 &              \\
              &           SDSS-r  &    73     & 17.67 & $1.087153\pm0.000007$ & 0.26 &              \\
              &           SDSS-i   &    72     & 18.03 & $1.087147\pm0.000005$ & 0.29 &              \\
              &           SDSS-z  &    70   & 18.35 & $1.087169\pm0.000021$ & 0.28 &              \\
        &   TESS      &    15746 &      & $1.089108\pm0.001332$ &      & \\
\noalign{\smallskip}
HS\,0713+3958 &CSS     &           434        & 16.61 & $0.782390\pm0.000017$ &              0.09 & DO UHE\\
              &           ATLAS-c              &           188        & 16.52 & $0.782404\pm0.000070$ & 0.08 &\\
              &           ATLAS-o             &           199        & 16.93 & $0.782537\pm0.000447$ & 0.08 &\\
              & ZTF-g   &       173     & 16.34 & $0.782351\pm0.000023$ & 0.11 &  \\
              & ZTF-r   &       193     & 16.89 & $0.782370\pm0.000023$ & 0.09 &  \\
              &           TESS      &           33045   &           & $0.782594\pm0.001509$ &        &\\
\noalign{\smallskip}                                                
HS\,0727+6003&            CSS        &           184        & 16.15 & $0.221410\pm0.000002$ &       0.13 &DO UHE \\
              &           ATLAS-c              &           121        & 16.08 & $0.221410\pm0.000003$ &       0.13 & \\
              &           ATLAS-o             &           135        & 16.51 & $0.221411\pm0.000060$ & 0.14 & \\
               & ZTF-g  &       202     & 15.90 & $0.221412\pm0.000002$ & 0.13 &  \\
               & ZTF-r  &       231     & 16.42 & $0.221409\pm0.000001$ & 0.13 &  \\
              &           BUSCA-U            &           399        &           & $0.221399\pm0.000003$ &       0.128 & \\
              &           BUSCA-B            &           495        &           & $0.221396\pm0.000010$ &       0.131 & \\
              &           BUSCA-R            &           493        &           & $0.221438\pm0.000015$ &       0.128 & \\
        &   TESS      &           17632   &           & $0.221453\pm0.000039$ &         & \\
\noalign{\smallskip}
J1059+4043 & ZTF-g  & 238  & 18.09 & $1.410591\pm0.000151$ & 0.08 & DOZ UHE \\
           & ZTF-r  & 229  & 18.68 & $1.410589\pm0.000154$ & 0.07 &  \\
\noalign{\smallskip}
J1215+1203&    CSS        & 441  & 18.20  & $0.601307\pm0.000011$ & 0.14 & DOZ UHE\\
               & ZTF-g  & 158  & 17.93  & $0.601319\pm0.000014$ & 0.14 &  \\
               & ZTF-r  & 173  & 18.51  & $0.601296\pm0.000035$ & 0.10 &  \\
\noalign{\smallskip}                            
J1257+4220 &   ATLAS-c              & 123  & 17.40  & $0.428993\pm0.000016$ &              0.18 &DA UHE                 \\
               & ZTF-g  & 287  & 17.24  & $0.428996\pm0.000006$ & 0.13 &  \\
               & ZTF-r  & 307  & 17.78  & $0.428993\pm0.000009$ & 0.11 &  \\
\noalign{\smallskip}
HS2027+0651 & ZTF-g  &  84  & 16.48  & $0.290784\pm0.000005$ & 0.06 &  DO UHE \\
            & ZTF-r  & 119  & 16.93  & $0.290782\pm0.000007$ & 0.05 &  \\
\noalign{\smallskip}
HS2115+1148 & ZTF-r  & 157  & 16.78  & $1.319665\pm0.000263$ & 0.02 &   DAO UHE\\
\hline
\noalign{\smallskip}
\end{tabular}
\end{table*}
 
\begin{table*}
\caption{Periods, mean magnitudes and amplitudes as derived from various light cures for all periodically variable white dwarfs
  showing only the He\,{\sc ii} line problem.}
\label{tab:periods2}
\begin{tabular}{l l r r c c l}
\hline\hline
\noalign{\smallskip}
Name & Band & Datapoints & Magnitude & $P$ & Amplitude & Comment \\
     &      &            &  [mag]    & [d] & [mag]     &         \\  
\hline
\noalign{\smallskip}
J0821+1739&    CSS        &           275        &           19.14    & $0.384835\pm0.000084$& 0.13 &DOZ UHE:      \\
              &           K2          &      2478           &           19.32    & $0.384878\pm0.000006$& &    \\
\noalign{\smallskip}
J1029+2540 &   CSS        & 470  & 17.11  & $0.282933\pm0.000016$  &  0.04 &DO UHE: \\
               & ZTF-g  & 130  & 16.85  & $0.282932\pm0.000007$  & 0.05 &  \\
               & ZTF-r  & 144  & 17.39  & $0.282926\pm0.000011$  & 0.04 &  \\
\noalign{\smallskip}
HE1314+0018    & TESS  &           13449   &           & $0.524170\pm0.001505$        & & DOZ \\
\noalign{\smallskip}
J1512+0651 & ZTF-r  & 119  & 17.56  & $0.226022\pm0.000010$ & 0.06 &  \\
\noalign{\smallskip}
HS1517+7403 & ZTF-g  & 259  & 16.42  & $1.091158\pm0.000057$ & 0.05 &DOZ  \\
            & ZTF-r  & 237  & 16.97  & $1.091142\pm0.000011$ & 0.04 &DO  \\
            & TESS   & 94063  &      & $1.091338\pm0.000278$ & & DOZ       \\
\noalign{\smallskip}
J1553+4832 & ZTF-g  & 1203 & 18.42  & $2.928482\pm0.000462$  & 0.05 &DO  \\
           & ZTF-r  & 1261 & 18.97  & $2.928408\pm0.000990$  & 0.04 &  \\
\hline
\noalign{\smallskip}
\end{tabular}
\end{table*}
 
\begin{table*}
\caption{Periods, mean magnitudes and amplitudes as derived from ZTF DR4 light
  cures for all periodically variable normal hot white dwarfs.}
\label{tab:periods3}
\begin{tabular}{l l r r c c l}
\hline\hline
\noalign{\smallskip}
Name & Band & Datapoints & Magnitude & $P$ & Amplitude & Comment \\
     &      &            &  [mag]    & [d] & [mag]     &         \\  
\hline
\noalign{\smallskip}
\object{KUV07523+4017}        & ZTF-g & 294  & 17.62 & $0.866092\pm0.000087$ & 0.05 & DOZ (PG\,1159) \\
\object{KUV07523+4017}        & ZTF-r & 443  & 18.12 & $0.866169\pm0.000098$ & 0.06 &  \\
\noalign{\smallskip}                         
\object{WDJ012828.99+385436.63} & ZTF-g & 154& 15.75 & $5.008217\pm0.001885$ & 0.06 & DA \\
\object{WDJ012828.99+385436.63} & ZTF-r & 192& 16.24 & $5.006654\pm0.002163$ & 0.05 &   \\
\noalign{\smallskip}                          
\object{WDJ031858.29+002325.66} & ZTF-g & 106& 18.44 & $3.527273\pm0.001443$ & 0.10 & DA \\
\noalign{\smallskip}                          
\object{WDJ055924.87+104140.41} & ZTF-r & 244& 17.49 & $0.570768\pm0.000058$ & 0.06 & DA (PN WeDe\,1) \\
\noalign{\smallskip}                          
\object{WDJ095125.94+530930.72} & ZTF-g & 222& 15.03 & $3.452674\pm0.000244$ & 0.20 & DA \\
\object{WDJ095125.94+530930.72} & ZTF-r & 450& 15.58 & $3.452675\pm0.000155$ & 0.20 &  \\
\noalign{\smallskip}                          
\object{WDJ112954.78+510000.26} & ZTF-g & 242& 17.52 & $2.895375\pm0.000366$ & 0.10 & DA \\
\object{WDJ112954.78+510000.26} & ZTF-r & 240& 18.04 & $2.895613\pm0.000691$ & 0.09 &  \\
\noalign{\smallskip}                          
\object{WDJ113905.98+663018.30} & ZTF-g & 330& 13.64 & $0.835974\pm0.000008$ & 0.18 & DAO+K7V \\
\object{WDJ113905.98+663018.30} & ZTF-r & 290& 13.59 & $0.835952\pm0.000005$ & 0.26 &  \\
\noalign{\smallskip}                          
\object{WDJ161613.10+252012.68} & ZTF-g & 138& 17.87 & $0.389031\pm0.000009$ & 0.09 & DA \\
\object{WDJ161613.10+252012.68} & ZTF-r & 154& 18.32 & $0.279841\pm0.000009$ & 0.07 &  \\
\noalign{\smallskip}                          
\object{WDJ162449.00+321702.00} & ZTF-r & 559& 16.26 & $1.095514\pm0.000069$ & 0.03 & DA+dM \\
\hline
\noalign{\smallskip}
\end{tabular}
\end{table*}

\newpage
 
\section{Figures}
 
\begin{figure*}[ht]
\centering
\includegraphics[width=\textwidth]{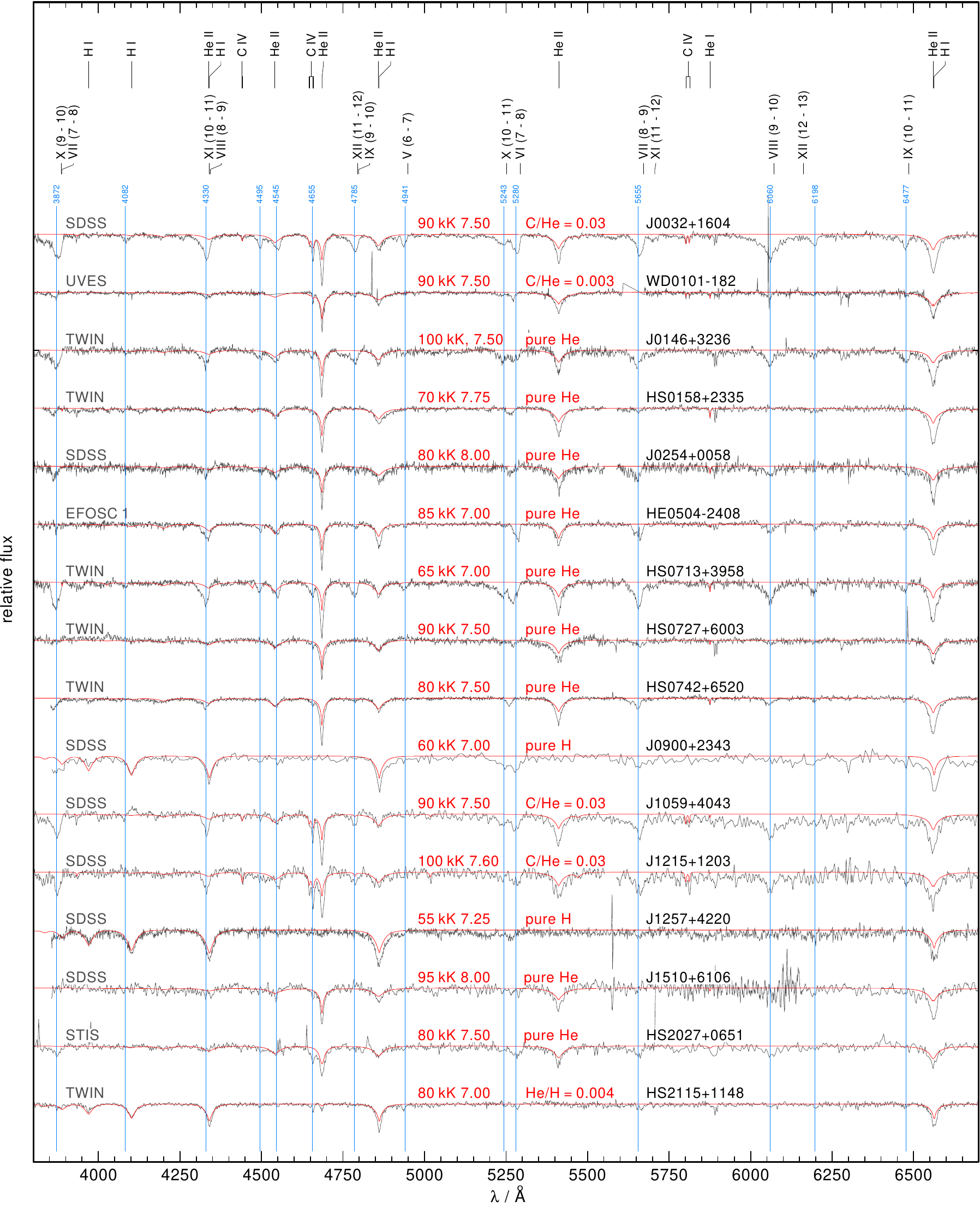}
\caption{Spectra of all known UHE white dwarfs. The positions of photospheric
  lines (H\,{\sc i}, He\,{\sc i}, He\,{\sc ii} and C\,{\sc iv}), $\alpha$ and
  $\beta$ transitions between Rydberg states ($n - n'$) of the ionization
  stages {\sc v}$-${\sc x}, and approximate line positions of the UHE features
  (blue) are marked. Overplotted in red are TMAP models and the effective
  temperatures, surface gravities, and chemical compositions (in mass
  fractions) - as determined in pervious works (see footnote of
  Table~\ref{tab:sample}) or here - are indicated.
  In gray the spectrograph used for the observation is indicated.}
\label{fig:spectraUHE}
\end{figure*}
 
\begin{figure*}[ht]
\centering
\includegraphics[width=\textwidth]{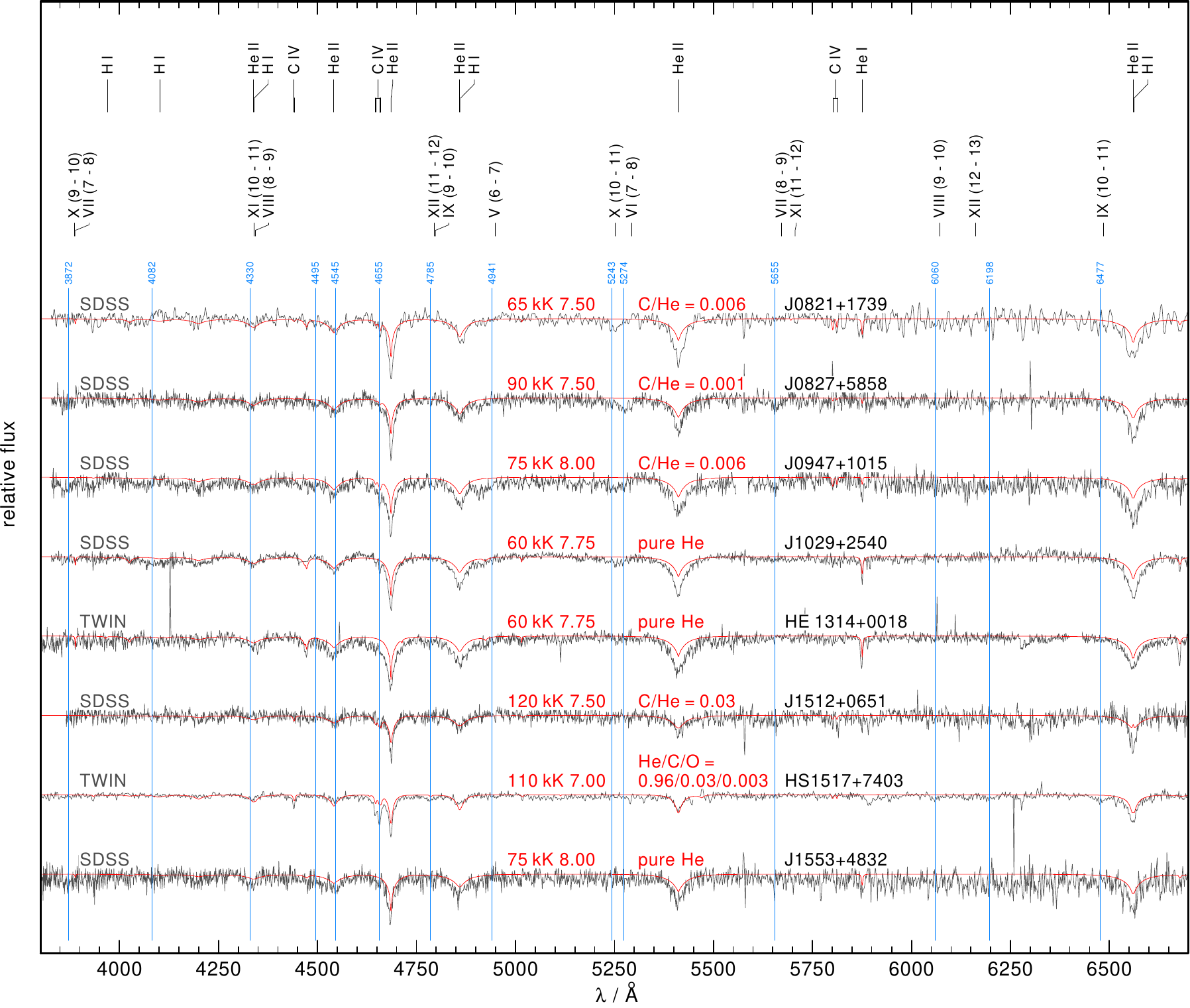}
\caption{Like \fg{fig:spectraUHE} for all known white dwarfs showing only the He\,II line
  problem but no UHE lines.}
\label{fig:spectraHeIIP}
\end{figure*}
 
\end{appendix}
 
\end{document}